\documentclass[twocolumn]{aastex631}

\usepackage{tablefootnote}

\begin{document}

\title{Type Ia Supernovae Can Arise from the Detonations of Both Stars in a Double Degenerate Binary}

\author[0000-0002-1184-0692]{Samuel J. Boos}
\affiliation{Department of Physics \& Astronomy, The University of Alabama, Tuscaloosa, AL, USA}

\author[0000-0002-9538-5948]{Dean M. Townsley}
\affiliation{Department of Physics \& Astronomy, The University of Alabama, Tuscaloosa, AL, USA}

\author[0000-0002-9632-6106]{Ken J. Shen}
\affiliation{Department of Astronomy and Theoretical Astrophysics Center, University of California, Berkeley, CA, USA}

\begin{abstract}
The precise origin of Type Ia supernovae (SNe Ia) is unknown despite their value to numerous areas in astronomy.
While it is a long-standing consensus that they arise from an explosion of a carbon/oxygen white dwarf, the exact progenitor configurations and explosion mechanisms that lead to SNe Ia are still debated.
One popular theory is the double detonation in which a helium layer, accreted from a binary companion, detonates on the surface of the primary star, leading to a converging shock-induced detonation of the underlying core. 
It has recently been seen in simulations that a helium-rich degenerate companion may undergo its own explosion triggered by the impact from the ejecta of the primary star.
We show 2D simulations that approximate a white dwarf undergoing a double detonation which triggers the explosion of the degenerate companion, leading to either a triple or quadruple detonation.
We also present the first multi-dimensional radiative transfer results from the triple and quadruple detonation scenario.
We find that within a range of mass configurations of the degenerate binary, the synthetic light curves and spectra of these events match observations as well as theoretical models of isolated double detonations do.
Notably, double and quadruple detonations that are spectrally similar and reach the same peak brightnesses have drastically different ejection masses and produce different amounts of Si- and Fe-group elements.
Further understanding of this scenario is needed in order to determine if at least some observed SNe Ia actually originate from two stars exploding.

\end{abstract}


\section{Introduction}
Type Ia supernovae (SNe Ia) are a crucial tool in astronomy as they, among other utilities, enabled the measurement of the local Hubble constant and led to the discovery of the accelerating expansion of the universe \citep{Riess_1998,Perlmutter_1999}.
Despite their use across a range of fields in astronomy, their precise origins are still unclear.
It is well-accepted that SNe Ia arise from a thermonuclear explosion of a carbon-oxygen white dwarf (WD), but the exact progenitor configuration and explosion mechanism is still debated.

There have been a wide variety of proposed scenarios which may lead a WD to explode as a SN Ia.
Since an isolated WD is essentially eternally stable, some influence from a companion is effectively required in order to trigger an explosion.
Both degenerate and non-degenerate companions have long been considered \citep{Whelan_1973, Iben_1984}, with varying degrees of predicted signatures and observational evidence \citep[see][for detailed reviews]{Maoz_2014,Liu_2023}.
In the typical single degenerate view where the companion is still hydrogen- or helium-burning, the WD accretes until it explodes around the Chandrasekhar-mass.
This scenario saw the bulk of the focus of the field for the first several decades, with the current successful models mildly reproducing some normal SNe Ia when undergoing some version of a ``delayed detonation'' \citep{Woosley_1986,Hoeflich_1995,Plewa_2004,Ropke_2007,Maeda_2010,Seitenzahl_2012,Sim_2013}.
Additionally, pure deflagrations of Chandrasekhar-mass white dwarfs might only be responsible for some subclasses of SNe Ia, including so-called ``Iax'' events \citep{Nomoto_1984,Jordan_2012,Kromer_2012,Fink_2013,Kromer_2015,Bulla_2020,Lach_2022}.

The double degenerate scenario, in which both stars are WDs, is somewhat more naturally consistent with observational characteristics of the majority of SNe Ia, including the elusiveness of the companion both pre- and post-explosion \citep{Li_2011,Bloom_2012,Kelly_2014,Kerzendorf_2018}.
Additionally, double degenerate systems are inherently very hydrogen-poor in most circumstances, which coincides with the defining feature of Type I supernovae, in addition to even late time observations of SNe Ia \citep{Tucker_2020}. 
In the double degenerate scenario, a SN Ia may be triggered via some merger event, either before or after the full disruption of the companion \citep{Pakmor_2010,Pakmor_2012,Kashyap_2015,Raskin_2018,Kashyap_2018,Neopane_2022}.
An increasingly promising theoretical explosion mechanism that may occur in the double degenerate scenario is the double detonation, which was first proposed by \citet{Nomoto_1982} \citep[see][for a history]{Townsley_2019}.
The double detonation may occur in a carbon/oxygen sub-Chandrasekhar white dwarf that has a helium shell, which may increase in shell mass by accretion from a helium-rich companion.
The helium shell detonates atop the primary WD and generates an inward-moving shock that converges within the core, igniting it and leading to a complete detonation of the WD \citep[see][for recent 3D simulations]{Tanikawa_2018,Gronow_2021,Pakmor_2022}.
This companion could theoretically be a non-degenerate star, but we focus on the prospect of a double degenerate system for this work in the context of supporting observational evidence.

A variety of studies have shown that the double detonation may indeed be a viable SN Ia channel, as the spectra and light curves around maximum light from explosion models can be satisfactory matches for some observed SNe Ia \citep{Sim_2010,Kromer_2010,Woosley_2011,Blondin_2017,Townsley_2019,shen_2021_nlte}.
Additionally, the breadth of many observed SNe Ia is also satisfied by theoretical models of double detonations \citep{Polin_2019,Shen_2021}.
This is possible due to the variable mass of the sub-Chandrasekhar progenitor and line of sight-dependence in double detonations which allows for a wide range of observational properties, including peak brightness and photospheric velocities \citep{Shen_2021,Collins_2022}.
The double detonation progenitor may also vary in composition and thickness of the surface helium layer which determine the production of high velocity, high-mass elements that can have a significant impact on the observables \citep{Woosley_2011,Polin_2019,Shen_2021}.
It is worth noting that while the double detonation has seen much success in the past decade, there remains some doubt about the scenario, particularly regarding the robustness of the core ignition prior to the complete disruption of the companion \citep{Moll_2013,Fenn_2016,Roy_2022}.

This wide array of variables is potentially a boon for the double detonation when it comes to explaining the span of most SNe Ia given the broad combinations of peak luminosities, decline rates, and spectral indicators that the candidate scenario can generate.
Because of this flexibility, the double detonation may be able to explain the scatter and outliers across the Philips relation.
For example, 2011fe and 2011by are a particular set of “twin” SNe that are extremely similar in their optical spectra and light curves, but differ markedly in the UV and have a relatively large peak magnitude difference of 0.33 \citep{Foley_2013,Graham_2015,Foley_2020}.
This specific contradiction to the standard SN Ia model has been suggested to be from progenitor metallicity differences \citep{Foley_2020}, though this is not conclusive and metallicity alone cannot explain all of the other peculiarities observed in the population of SNe Ia.
It is possible that some of these may be solved by the right combination of progenitor parameters and observed line of sight in the double detonation scenario.
That is, one might be able to change multiple parameters of the underlying system in such a way as to keep something as specific as the maximum light optical spectrum fixed while other features such as the UV and maximum brightness vary.

This flexibility, however, can present a challenge when trying to interpret observables within the traditional one-parameter family model of SNe Ia. 
To this end, these variations are often characterized as ``scatter'' about some average one-parameter family. 
Ultimately, a fully functional model of SNe Ia should resolve this distinction, providing physical explanations for both the general one-parameter variation as well as the departures from it.
There is much work to be done in this respect, as the capability to simulate the double detonation scenario continues to mature.
Additionally, the need for costly non-local thermodynamic equilibrium (non-LTE) radiative transfer calculations has become more evident, as some observational features, of both the niche and ubiquitous variety, require such calculations \citep{Boyle_2017,shen_2021_nlte,Collins_2023}.
This is a significant issue for many supernova models, but it is especially burdensome for the double detonation scenario given its inherent multidimensionality.

Another possible variable affecting the observables of double detonation events is the fate of the donor companion.
Given that the double detonation occurs in a tight, likely double degenerate, binary and the primary WD is completely detonated, the donor may be expected to be ejected as a high velocity runaway.
While evidence of such a surviving star long eluded observation, a handful of candidates have been identified in recent years using \emph{Gaia} data \citep{Shen_2018,ElBadry_2023}.
Alternatively, it may be possible that the donor WD can also be destroyed in the event.
\citet{Papish_2015} first showed a simulation where a helium companion detonates following the impact from a double detonation, i.e. a ``triple detonation''.
A pair of recent computational studies \citep{Tanikawa_2019,Pakmor_2022} have examined the scenario in which a helium-shelled carbon/oxygen companion undergoes its own double detonation following the initial double detonation, i.e. a ``quadruple detonation''.
These 3-dimensional studies show the mechanics of how the core detonation of the primary triggers a delayed detonation of the donor’s helium shell, leading to a second double detonation.

The predicted observables from these unique detonations are scant, however.
\citet{Pakmor_2022} presented 1-dimensional radiative transfer results for their sole 1.05 + 0.7 $M_{\odot}$ quadruple detonation model and found that the additional mass and burning yields from the companion detonation had surprisingly little impact on the observables, albeit calculated with an angle-averaged ejecta profile for a low-mass and low-Fe-group-generating companion.
This result begs the question of how the quadruple detonation scenario fares across various lines of sight for a range of binary mass configurations, particularly those with a relatively high-mass companion.

Some rare objects display overluminous brightnesses so high that they have been labeled ``super-Chandrasekhar'' due to the inferred ejected mass \citep{Howell_2006,Ashall_2021}.
Attempts to explain these events have included the explosion of a super-Chandrasekhar WD \citep{Hachinger_2012} and violent merger of a high-mass degenerate binary \citep{Pakmor_2012}.
Having both stars in a high-mass binary explode presents an interesting candidate for these objects.
The number of WD binaries with total masses much above the Chandrasekhar limit are expected to be a fraction of their lower mass counterparts \citep{nelemans_2001}, so these events would likely be rare.
It is possible, however, that if a few of these events have been observed, they may have appeared normal enough amongst the breadth of SNe Ia to be automatically assumed to be from a single star detonation.
This additional degree of freedom is interesting from the standpoint of observed diversity among SNe Ia.
In the double detonation scenario, both the $^{56}$Ni mass and the total ejecta mass are determined by the primary WD mass, and therefore are directly tied.
The two star scenario breaks this by allowing explosions with the same $^{56}$Ni mass to have different total ejecta masses depending on the mass of the secondary.

While the conclusion from \citet{Pakmor_2022} is extremely interesting, multidimensional analysis across a wider range of progenitor configurations is necessary to evaluate this scenario further. 
To that end, this work simulates a number of 2D two star explosion models and presents their multi-dimensional synthetic observables.
The progenitors used in this work are bare WDs (for the purpose of reduced computational complexity), but approximate the dynamics of triple and quadruple detonations that may otherwise occur when helium shells are considered.
We refer to these models as triple and quadruple detonations in this work, despite the exclusion of helium in our calculations.
Our models consist of a number of binary mass configurations, including those that produce significant amounts of radioactive material in the companion detonation.

We describe our computational setup and choice of progenitor systems in Section \ref{sec:methods}.
The multi-dimensional light curves and spectra from these two star explosion models are shown in Section \ref{sec:results} and compared to both our previous single star double detonation models and observed SNe Ia. 
We then discuss the overview of our results in the context of observed SNe Ia in Section \ref{sec:discussion} before summarizing this work in Section \ref{sec:conclusions}.

\section{Methods}\label{sec:methods}

We use the multiphysics code \texttt{FLASH} \citep{Fryxelletal2000} to simulate the detonations in this work, in a similar manner as \citet{Boos_2021}.
We use an adaptive mesh with a minimum cell size of 8 km.
We use the Helmholtz EOS and aprox13 nuclear network, along with a burning limiter.
50,000 equal-mass tracer particles are distributed by density within each progenitor.
The temperature and density histories from these particles are later used in nucleosynthetic post-processing using \texttt{MESA} and a 205-nuclide network.
The significant modifications to our setup from previous work, which involve the change of nuclear network, exclusion of helium shells, and inclusion of the companion WD, are described in Sections \ref{subsec:network_and_shell} -- \ref{subsec:binarysetup}.

To generate the spectra from the detonation simulations, we use the Monte Carlo radiative transfer code \texttt{Sedona} \citep{Kasen_2006} under the assumption of LTE.
The manner of our radiative transfer calculations are unchanged from that described in \citet{Shen_2021}, other than a three-fold reduction of particles and using ejecta out to maximum velocities of $4.5 \times 10^9$ cm s$^{-1}$, rather than the original $3 \times 10^9$ cm s$^{-1}$.
This increase in velocity domain is important for the prediction of the UV region of the spectrum.

\subsection{Nuclear Reaction Network Choice and the Helium Shell}\label{subsec:network_and_shell}

In this work, we choose to use the nuclear network aprox13 \citep{Timmes_2000,Fryxelletal2000}, which uses 13 isotopes, for our detonation simulations over the 55 isotope network previously used in \citet{Townsley_2019} and \citet{Boos_2021} to significantly improve the speed of these calculations.
While this network reduces the accuracy of the energy release somewhat, our temperature-density history post-processing technique \citep{Boos_2021} generates similar results between the two networks used in the detonation simulation.
The reduction of the nuclear network prevents a self-sustaining detonation in thin helium shells (see \citealt{Moore_2013} and \citealt{Shen_and_Moore_2014} for detailed work on helium shell detonations).
Since we are unable to simulate a thin helium shell detonation in this setup and thick helium shells have been shown to generate observables that can only reproduce a fraction of peculiar events \citep{Woosley_2011,Polin_2019,Shen_2021,ChangLiu_2023,Padilla_Gonzalez_2023}, we choose to exclude the helium shell in these simulations.

To evaluate our additional approximations (the lack of helium shell and reduced nuclear network) of our setup, we conduct a detonation of an isolated, non-shelled WD in our new setup to compare to the true double detonations of \citet{Boos_2021}.
In this comparison simulation, we detonate a 1.00 $M_{\odot}$ C/O WD by way of an artificial hotspot within the star.
The hotspot is placed where the shock from the helium shell detonation converged in the counterpart model in \citet{Boos_2021}.

\begin{table}
\centering
\caption{Approximated yields}
\begin{tabular}{c|c|c|c|c|c|c}
M$_{core}$ & N$_{iso}$\footnote{Number of isotopes considered in hydronuclear simulation} & Shell\footnote{Presence and detonation of helium shell} & $^{12}$C & $^{28}$Si & $^{40}$Ca & $^{56}$Ni \\
($M_{\odot}$) & & & ($M_{\odot}$) & ($M_{\odot}$) & ($M_{\odot}$) & ($M_{\odot}$) \\
\hline
\hline
1.00 & 13 & No & 1.9$\times 10^{-3}$ & 0.17 & 1.6$\times 10^{-2}$ & 0.55 \\
1.00 & 55 & Yes & 2.8$\times 10^{-3}$ & 0.18 & 1.9$\times 10^{-2}$ &  0.53 \\
\end{tabular}
\label{table:shell_vs_no_shell_yields}
\end{table}

To determine the nuclear burning and observational effects from the lack of helium shell and reduced nuclear network in the simulation, we compare this non-shelled model with the 1.00 $M_{\odot}$ core, 0.02 $M_{\odot}$ helium shell model from \citet{Boos_2021}.
These models have the same C/O mass and similar central densities.
The differences in post-processed yields between the core detonations of these models are shown in Table \ref{table:shell_vs_no_shell_yields}.
The 205-nuclide network used in the post-processing is the same for both models.
We find that the core nucleosynthetic yields are fairly similar between the two cases.
Around 0.02 $M_{\odot}$ more $^{56}$Ni is produced in the shelled progenitor using a larger network in the hydrodynamic simulation.
This slight increase in burning may be attributable to the minor density enhancement of the core from the helium detonation or the more accurate nuclear network.

\begin{figure}
    \centering    \includegraphics[width=0.5\textwidth]{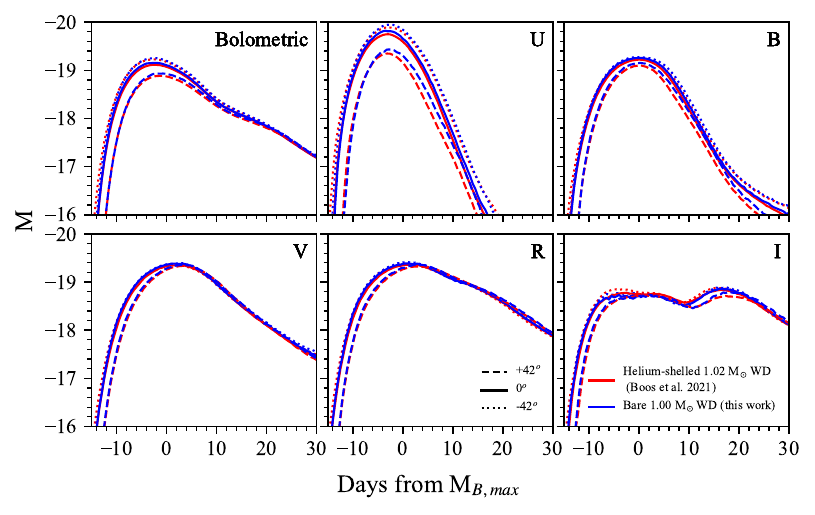}
    \caption{Multiband light curves from the detonation of an isolated 1.00 $M_{\odot}$ WD, compared with that of a double detonation of a thin helium shell 1.02 $M_{\odot}$ WD \citep{Boos_2021,Shen_2021}.
    Three lines of sight from each model are shown, where the dotted, solid, and dashed lines represent the model as observed from a southern, equatorial, and northern line of sight, respectively.
    }
    \label{fig:isolated_light_curves}
\end{figure}

Figure \ref{fig:isolated_light_curves} shows the light curves of these two models for three lines of sight.
Despite the lack of a helium shell and use of a less complete nuclear network, the light curves are extremely similar between the two models.
The largest deviation is seen in the post peak decline at the northern line of sight in U-band, but for the most part, the light curves are very nearly the same.
We note that ``north'' in this work is defined as the positive z-direction of our simulations.

\begin{figure}
    \centering    \includegraphics[width=0.5\textwidth]{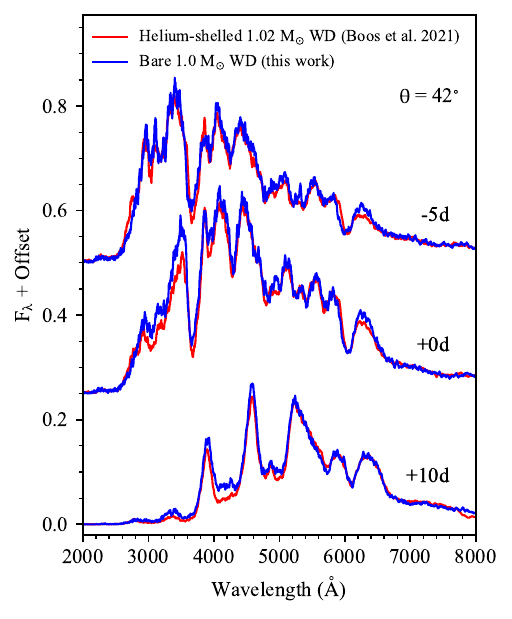}
    \caption{Spectral comparison between the 1.02 $M_{\odot}$ shelled double detonation model from \citet{Boos_2021} and a bare 1.00 $M_{\odot}$ C/O WD 42\textsuperscript{o} above the equatorial plane, the line of sight at which these models disagree the most.
    Each pair of spectra are at the labeled time relative to their respective B-band maxima.
    }
    \label{fig:isolated_spectra}
\end{figure}

Figure \ref{fig:isolated_spectra} shows a spectral comparison of these two models at various times around the peak B band time at the line of sight where they disagree the most (42\textsuperscript{o} north of the equatorial plane).
The spectra are very similar at each time, though the UV portion is slightly enhanced at peak B-band time and later in the bare C/O WD case, likely due to the lack of line blanketing from shell ashes.
Additionally, there is a lack of late-time suppression around 4100 {\AA} that is not observed at lines of sight south of those shown in Figure \ref{fig:isolated_spectra}.

In summary, there is remarkably little difference in the observables between explosions of two progenitors that are nearly identical, save for the presence of a thin helium shell.
Thus, we determine that it is practical to model thin shell double detonations without the shell material or detonation, especially for the approximated evaluation of the two star explosion scenario performed in this work.

\pagebreak 

\subsection{Setup of the binary}\label{subsec:binarysetup}

\begin{table*}
\centering
\caption{Model Details}
\begin{tabular}{cc|c|c|c|c|c|c|c}
M$_P$\footnote{Primary mass} & M$_C$\footnote{Companion mass} & Companion & Companion & R$_P$\footnote{Primary radius} & R$_C$\footnote{Companion radius} & Separation\footnote{Distance between primary origin and companion center at time of companion ignition} & $\rho_{C,0}$\footnote{Pre-interaction density of the companion at its ignition point} & Core Ignition Delay \\ 
(M$_{\odot}$) & (M$_{\odot}$) & Composition & Ignition Mode & ($10^{3}$ km) & ($10^{3}$ km) & ($10^{3}$ km) & ($10^{6}$ g cc$^{-1}$) & (s) \\ \hline \hline
0.85 & 0.80 & C/O & Interior & 6.78 & 7.20 & 20.8 & 6.02 & 3.65 \\ 
1.00 & 0.70 & C/O & Interior & 5.58 & 8.09 & 21.8 & 5.37 & 3.35 \\ 
1.00 & 0.70 & C/O & None & 5.58 & 8.09 & 21.8 & --- & --- \\ 
1.00 & 0.90 & C/O & Interior & 5.58 & 6.38 & 18.1 & 6.58 & 2.90 \\ 
1.00 & 0.40 & He & Direct & 5.58 & 14.06 & 38.2 & 0.28 & 2.40 \\ 
1.10 & 1.00 & C/O & Interior & 4.77 & 5.58 & 15.7 & 8.50 & 2.90 \\ 
1.10 & 1.00 & C/O & Direct & 4.77 & 5.58 & 15.6 & 1.28 & 1.10 \\ 
\end{tabular}
\label{table:model_details}
\end{table*}

The details for each of our two star simulations are summarized in Table \ref{table:model_details}.
At initialization, the primary star is placed at the origin of the grid as it was in \citet{Boos_2021}.
The secondary is placed offset from the primary in the positive z-direction, with the central axis of both stars aligned with the axis of symmetry for the simulation.
At the time of disruption, the system separation should be such that the donor fills its Roche lobe.
Since the binary has no angular momentum in these 2D simulations, a companion initialized at the Roche limit at rest will gain velocity and move towards the primary within the few seconds it takes for the first detonation to influence the companion.
So, we initialize the companion WD at a separation slightly below the Roche lobe radius with a velocity ($\sim$ several 10$^7$ cm s$^{-1}$) away from the primary such that the companion is at the Roche limit with roughly no velocity when the impact shock from the primary detonation is propagating through it.
The low-density, domain-filling material outside of the stars (referred to as ``fluff''), is slightly changed from the scheme detailed in \citet{Boos_2021}.
The fluff now has a uniform density within a radius that encloses both progenitors, outside of which it declines in a log-linear manner.

Since the helium detonations that generate the converging-shock ignition of the C/O core are not considered in this work, we use artificial hotspots to ignite all of the primary WDs, in addition to most of the companions.
The location of these hotspots are informed by the shock convergence points found in previous works that considered helium shells for both double and quadruple detonations.
This setup preserves the off-centered interior ignition and asymmetric progression of the core detonation that is characteristic to the double and quadruple detonation.
The dynamics leading to each of the ignitions in this scenario is complex in reality, and may be particularly so for the companion helium detonation and subsequent core ignition due to the influence from the accretion and primary WD explosion.
However, these dynamics are not the focus of this study; rather we seek to understand how these events may be observed assuming that the two star detonation in this scheme is plausible.

The circular hotspots used in this work are between 200 and 400 km in radius and have a temperature profile that peaks at $2 \times 10^{9}$ K in the center and linearly declines to roughly the local star temperature.
We note that the critical hotspot sizes for each of these ignitions were not rigorously investigated in this work (see \citealt{Seitenzahl_2009} for critical sizes of this hotspot profile) given the relatively modest resolution and uncertain location and condition of the companion ignition.
These hotspots are chosen slightly large to ensure ignition, so this study should be viewed as an evaluation of whether such explosions are observationally viable under the assumption that ignition occurs.
We leave a thorough exploration into the initiation of the companion core detonation in particular to future works.

At initialization for each of our simulations, a hotspot is placed 100 km from the symmetry axis in the southern hemisphere of the primary, where the shock generated from a helium detonation ignited at the northern pole would otherwise converge and ignite the core (see Figures 2 and 3 of \citet{Boos_2021} for a natural C/O core ignition).
In some of our simulations, this ultimately leads to an ignition of the companion WD along its southern edge where the impact from the primary ejecta is the strongest, which we label as ``direct'' ignition cases.
If the companion is not directly ignited, we place a hotspot along the symmetry axis in the northern hemisphere of the interior region of the companion core, corresponding to the focus location of the shock created by a helium shell detonation that would be ignited at the southern pole, nearest the primary.
Based on \citet{Pakmor_2022} and a demonstration simulation with our setup using high-mass helium shells, this second WD ignition would be driven by the combination of shocks from both the companion helium detonation and primary WD detonation.
Since the progenitors in this approximated work are bare of helium shells, only the shock from the primary WD detonation exists in our simulations.
So, we roughly place the second hotspot in the companion at the position where the shock from a helium detonation would converge ($\sim1-3 \times 10^{3}$ km from the companion center) at the time when the primary ejecta shock intercepts this location.
We refer to these models as ``interior'' ignition cases, which approximate quadruple detonations.

We present two cases that undergo a direct ignition of the companion.
The first is a model with a 0.40 $M_{\odot}$ helium companion that detonates following the detonation of a 1.00 $M_{\odot}$ carbon/oxygen primary.
In this case, the helium companion ignites naturally when it is first impacted by the primary ejecta, producing the triple detonation similar to \citet{Papish_2015,Tanikawa_2019}.
Like \citet{Papish_2015,Tanikawa_2019}, we also find that the direct ignition of the helium companion is sensitive to the separation of the stars.
We find a separation threshold of $38.2 \times 10^8$ cm for our 1.00 + 0.40 $M_{\odot}$ system, which is several $10^8$ cm lower than the Roche lobe radius for this system.
The helium companion ignites at a region of the star that is shocked to a density and temperature of $8.02 \times 10^{5}$ g cm$^{-3}$ and $1.20 \times 10^{9}$ K, from an original pre-shock state of $2.75 \times 10^{5}$ g cm$^{-3}$ and $3.00 \times 10^{7}$ K.
This companion is made up of pure helium in the grid simulation, but is post-processed with 0.009 $^{14}$N (as in \citealt{Boos_2021}) and elements above Z $= 8$ are scaled to solar metallicity.

The other direct ignition case is an alternate version of the 1.10 + 1.00 $M_{\odot}$ progenitor system that has a slightly smaller separation than its interior  ignition counterpart, resulting in a direct ignition at the impact point on the near edge of the carbon/oxygen companion.
This simulation has similar dynamics to the helium companion case, however in reality this carbon/oxygen companion may have some helium remaining on its surface when the system is first ignited which may suppress the ability of the primary ejecta impact to directly ignite the companion.
For the purpose of this work, we wish to characterize both the interior and direct ignition possibilities.
As such, we also consider this direct ignition model as an approximated triple detonation, under the supposition that the companion helium that is not considered in this work would have relatively little effect on the delay or occurrence of the ignition of the companion core.

Additionally, we show the results of a 1.00 + 0.70 $M_{\odot}$ binary where the secondary is not ignited and survives the impact from the primary detonation, which gives insight on how the surviving companion may affect the ejecta and observables of a single double detonation.
In this simulation, the companion remains near the origin throughout the simulation.
In order to perform radiative transfer calculations, we need to remove the bound material from the end-of-simulation ejecta.
To that end, we cutoff the ejecta inward of 1,600 km s$^{-1}$, removing all of the bound material.
We then calculate the mass and average composition of the unbound ejecta between 1,600 and 2,500 km s$^{-1}$, and uniformly set the density and composition below 2,500 km s$^{-1}$ such that the nucleosynthetic yields and mass of the unbound ejecta is preserved, filling the void created by the removal of the bound material.
We note that this method is likely sufficient for maximum light observables (which are the focus of this work) because this region of the ejecta is located far beneath the photosphere during this phase.
However, a more careful treatment of the bound remnant may be necessary for accurate production of late time observables.

\section{Results} \label{sec:results}
We present here the results of our simulations.
In Section \ref{subsec:detonations_and_ejecta}, we detail the nuclear-hydrodynamic dynamics of our two star explosion models as well as the resulting ejecta structure.
We present the nucleosynethic yields and synthetic observables for our models in Sections \ref{subsec:yields} and \ref{subsec:synthetic_observables}, respectively.
Finally, we compare  in Section \ref{subsec:observational_correlations} a variety of observed correlations of SNe Ia to that of our complete suite of double, triple, and quadruple detonation models.

\subsection{Detonation Dynamics and Ejecta Morphology}
\label{subsec:detonations_and_ejecta}

\begin{figure}
    \centering
    \includegraphics[width=0.5\textwidth]{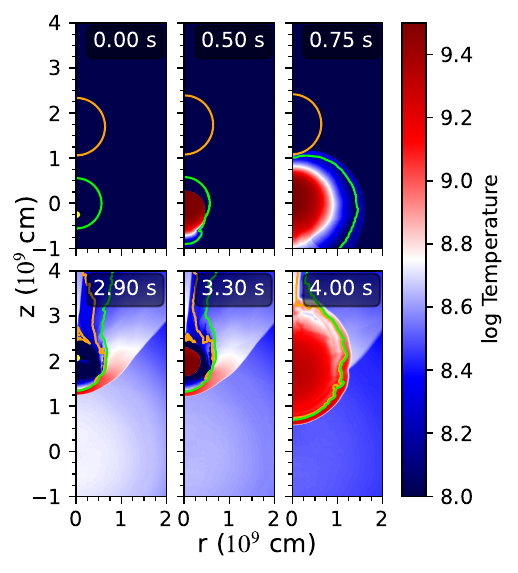}
    \caption{Temperature of the 1.00 + 0.90 $M_{\odot}$ quadruple detonation model during the detonation phase.
    The primary and secondary cores are ignited by hand at 0.00 and 2.90 seconds, respectively, and their ignition locations near the symmetry axis are given by a yellow semicircle (not indicative of artificial hotspot size).
    The green and orange lines are the 90\% contours for primary and secondary star material, respectively. 
    }
    \label{fig:walkthrough}
\end{figure}

A demonstration of a quadruple detonation simulation from this work is shown in Figure \ref{fig:walkthrough}.
The first frame shows the primary and secondary WDs at initialization, when a hotspot is placed in the southern hemisphere of the primary star.
The detonation is seen progressing through the primary in the second frame.
The shock from this detonation first reaches the southern surface of the companion 0.75 seconds after the primary is ignited.
The fourth frame in Figure \ref{fig:walkthrough} shows the point at which the companion is manually ignited in the northern hemisphere.
The detonation is then seen propagating in the companion in the fifth frame, before it is completed by around 4 seconds at which point the ejecta from the two explosions begin to evolve into a homologous state.

This sequence occurs similarly in our quadruple detonation models, albeit with slightly different delays between detonations (around 3 to 3.5 seconds) due to the varying mass configurations.
In two of our runs, we find that the companion ignites directly at its southern edge when it is initially impacted by the primary ejecta, producing a triple detonation.
This occurs in the case with a 0.40 $M_{\odot}$ helium WD and in an alternate version of the 1.10 + 1.00 $M_{\odot}$ model where we place the companion closer to the primary than in the counterpart model that undergoes the aforementioned interior ignition at the Roche limit.

\begin{figure*}
    \centering
    \includegraphics[width=1.0\textwidth]{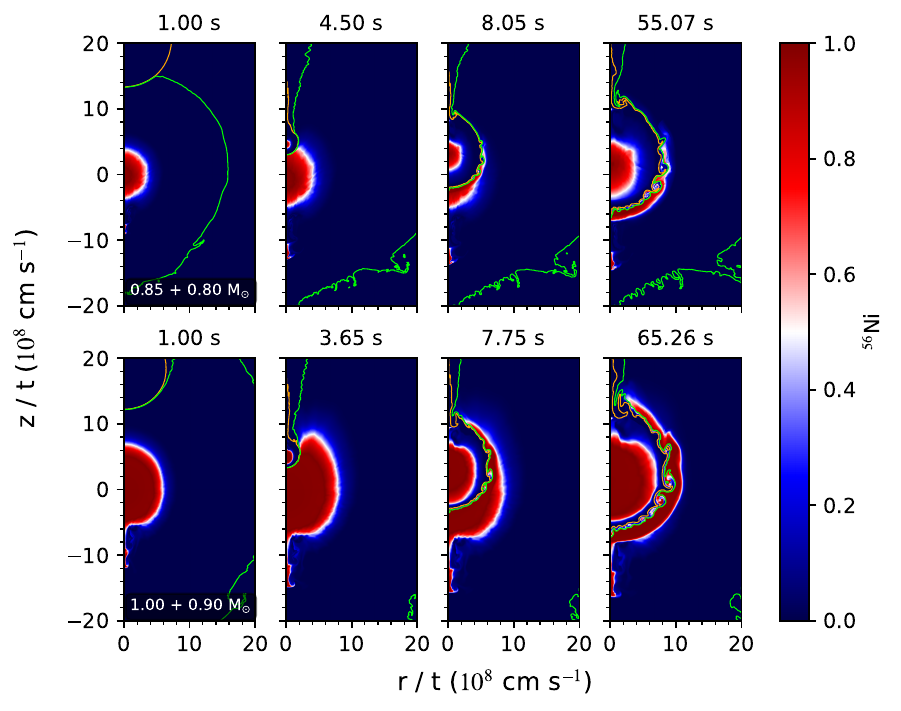}
    \caption{Formation of the $^{56}$Ni distribution in the 0.85 + 0.80 $M_{\odot}$ (top) and 1.00 + 0.90 $M_{\odot}$ (bottom) quadruple detonation models. 
    The green and orange lines are the 90\% contours for primary and secondary star material, respectively. 
    From left to right, the column pairs represent the simulations one second after the primary is ignited, when the companion detonation is complete, a few seconds into the ejecta expansion phase, and when the ejecta is in homologous expansion.
    The axes represent distance from the initial position of the primary star, scaled by the time after first ignition.
    The values of the axes in the final frame represent the asymptotic ejecta velocities.
    }
    \label{fig:primcmpn_and_ni56}
\end{figure*}

The development of the quadruple detonation ejecta can be seen in Figure \ref{fig:primcmpn_and_ni56}.
Before the companion explodes, the shape of the primary star ejecta and distribution of $^{56}$Ni looks similar to that seen in the core detonations of the isolated models of \citet{Boos_2021}, aside from the ``shadow'' generated by the companion blocking expanding primary ashes at northern angles (see \citealt{Papish_2015,Tanikawa_2018,Ferrand_2022}).
This distribution is disrupted a few seconds later when the companion ejecta expands into the surrounding primary ashes, compressing a substantial amount of the innermost primary ashes into a shell-like structure.
By the time the ejecta becomes homologous, it has a structure that is a notable departure from that of the isolated double detonation.
Specifically, the outermost ashes of the companion detonation are reduced in velocity as the ejecta expands into relatively dense primary star material.
More importantly, the $^{56}$Ni generated in the primary detonation is pushed out to at least 5,000 km/s in both models shown in Figure \ref{fig:primcmpn_and_ni56}.
This is a notable change from isolated double detonation models given the $^{56}$Ni distribution typically has velocities of at most $\sim$5,000 km/s, depending on progenitor mass.
This two zone structure of radioactive material is reminiscent of thick helium shell double detonations of single stars, where the detonation in high density helium shells produces significant amounts of $^{56}$Ni which end up at higher velocities.
Such thick helium shell double detonations have extensively been shown to be a poor candidate for normal SNe Ia \citep{Woosley_2011,Polin_2019,Shen_2021}.
It is shown in Section \ref{subsubsec:snia_like_twostar}, however, that this is not the case with the quadruple detonation models, likely due to the distribution of predominately Si group elements at high velocity.

The added ejecta formation dynamics from the second star explosion inject an additional layer of asymmetry into the double detonation model.
This is due to the position of the companion being modestly off-centered from the origin of the the primary ejecta by the time it starts expanding.
This results in a significant reduction of inner primary ashes at northern angles, which is most clearly seen in the top-right frame of Figure \ref{fig:primcmpn_and_ni56}, where hardly any $^{56}$Ni from the primary is northward of the equatorial plane in the 0.85 + 0.80 $M_{\odot}$ model.
Thus, observers viewing this event from opposite poles may see drastically different ejecta structures.
Comparing this to the 1.00 + 0.90 $M_{\odot}$ model (bottom-right frame of Figure \ref{fig:primcmpn_and_ni56}), it is shown that this asymmetrizing feature of the quadruple detonation is also dependent on binary mass configuration, as $^{56}$Ni from the primary detonation is present, albeit with varying amounts, at all lines of sight.

\begin{figure*}
    \centering
    \includegraphics[width=1.0\textwidth]{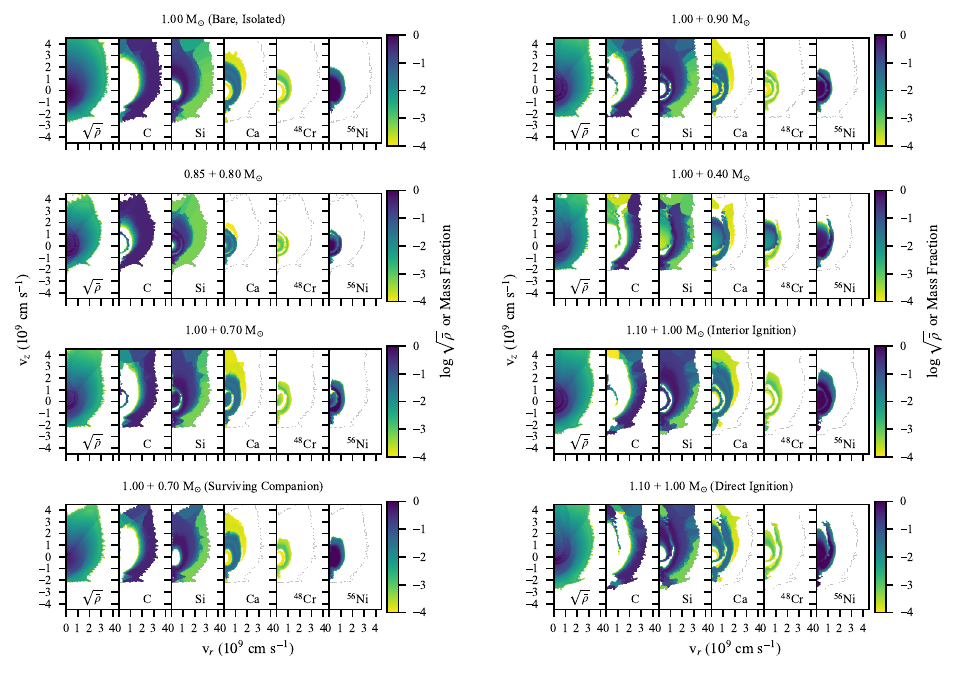}
    \caption{Density and composition maps of each of our models in velocity space.
    The qualitative representation of the density is the logarithm of the square root of the density scaled by the maximum value in each respective model.
    The composition shown is the logarithmic mass fraction of the post-processed data.
    Logarithmic values below -4 are colored white, along with material that does not originate from either star (i.e. ``fluff'').
    The top-left case shown is an isolated WD model, detailed in Section \ref{subsec:network_and_shell}.
    The remaining seven models involve a binary system, with all but the bottom-left case having both stars explode.
    }
    \label{fig:all_ejecta}
\end{figure*}

The 2D ejecta in velocity space for each of our two star models, in addition to an isolated WD model, are shown in Figure \ref{fig:all_ejecta}.
In general, we observe a relatively consistent ejecta structure in the quadruple explosion models where the companion ashes are embedded within the ashes of the primary, similar to what is seen in \citet{Tanikawa_2019} and \citet{Pakmor_2022}.
This is most clearly demonstrated in the 2D profiles of $^{28}$Si, which is generated in the outer region of each of the massive progenitors and throughout the $0.70$ $M_{\odot}$ ashes.
In the isolated WD explosion, one clear band of $^{28}$Si can be observed, peaking between 10,000 and 20,000 km s$^{-1}$, depending on polar angle.
In the two star explosion cases, there are now two prominent bands of $^{28}$Si, each belonging to one of the exploding stars.
The velocity extent of these two concentrations are significantly different due to the suppressed expansion of the companion ashes imposed by the surrounding primary ejecta.
A similar structure can also be seen in the $^{56}$Ni of our two star explosions models where the $^{56}$Ni from the primary detonation is found outside that of the companion, separated by a relatively narrow layer of predominantly  intermediate mass elements that originates from the companion detonation.
While the 3D simulations of similar binary star explosions in \citet{Tanikawa_2019,Pakmor_2022} show very similar ejecta stratification compared to our models, we find that our ejecta show less prominent asymmetrization, possibly due to multidimensional effects.

In the pair of triple detonation cases where the companion star is directly ignited (1.00 + 0.40 $M_{\odot}$ and 1.10 + 1.00 $M_{\odot}$ (direct ignition)), rather than via a delayed interior ignition, the ejecta structures are slightly different.
Due to the smaller delay between star ignitions, these two models yield significantly increased velocities of the companion ashes at northern latitudes.
A similar effect on the ejecta is seen in the triple detonation case from \citet{Tanikawa_2019}.

We note that a more precise prediction of the final state of the ejecta would demand the inclusion of the shell detonations.
Not only will these shell detonations add two additional layers of Si-group material in thin-shell cases \citep{Polin_2019,Boos_2021}, but they may also affect the timing and location of the core ignition in the companion.
The placement of the artificial hotspots used to ignite the companions in this paper were influenced by ignition timings and locations of the core detonations in similar, but shelled, progenitors in \citet{Boos_2021}, but it is unclear exactly how the interaction between the converging shock from the helium shell detonation might interact with the northerly-moving shock from the primary detonation.
For example, the companion core detonation may trigger earlier than expected, allowing for more companion ejecta to expand to higher velocities.

\subsection{Yields}
\label{subsec:yields}

\begin{table*}
\centering
\caption{Binary Explosion Yields}
\begin{tabular}{cc|c|cc|cc|cc|cc}
M$_P$\footnote{Primary mass} & M$_C$\footnote{Companion mass} & Companion & $^{12}$C$_{P}$ & $^{12}$C$_{C}$ & $^{28}$Si$_{P}$ & $^{28}$Si$_{C}$ & $^{40}$Ca$_{P}$ & $^{40}$Ca$_{C}$ & $^{56}$Ni$_{P}$ & $^{56}$Ni$_{C}$ \\
(M$_{\odot}$) & (M$_{\odot}$) & Ignition Mode & (M$_{\odot}$) & (M$_{\odot}$) & (M$_{\odot}$) & (M$_{\odot}$) & (M$_{\odot}$) & (M$_{\odot}$) & (M$_{\odot}$) & (M$_{\odot}$) \\
\hline
\hline
0.85 & 0.80 & Interior & 1.3$\times 10^{-2}$ & 9.0$\times 10^{-3}$ & 0.25 & 0.24 & 2.0$\times 10^{-2}$ & 1.9$\times 10^{-2}$ & 0.19 & 0.15 \\
1.00 & 0.70 & Interior & 1.8$\times 10^{-3}$ & 3.3$\times 10^{-2}$ & 0.17 & 0.22 & 1.6$\times 10^{-2}$ & 1.2$\times 10^{-2}$ & 0.55 & 2.1$\times 10^{-2}$ \\
1.00 & 0.70 & None & 1.7$\times 10^{-3}$ & - & 0.17 & - & 1.6$\times 10^{-2}$ & - & 0.55 & - \\
1.00 & 0.90 & Interior & 1.9$\times 10^{-3}$ & 3.5$\times 10^{-3}$ & 0.17 & 0.2 & 1.6$\times 10^{-2}$ & 1.8$\times 10^{-2}$ & 0.55 & 0.39 \\
1.00 & 0.40 & Direct & 1.9$\times 10^{-3}$ & 1.6$\times 10^{-4}$ & 0.17 & 4.0$\times 10^{-4}$ & 1.6$\times 10^{-2}$ & 2.2$\times 10^{-2}$ & 0.55 & 0.11 \\
\hline
1.10 & 1.00 & Interior & 1.4$\times 10^{-3}$ & 1.1$\times 10^{-3}$ & 0.11 & 0.13 & 1.2$\times 10^{-2}$ & 1.4$\times 10^{-2}$ & 0.79 & 0.61 \\
1.10 & 1.00 & Direct & 8.4$\times 10^{-4}$ & 3.8$\times 10^{-4}$ & 0.13 & 0.15 & 1.2$\times 10^{-2}$ & 1.4$\times 10^{-2}$ & 0.79 & 0.55 \\
\end{tabular}
\label{table:yields}
\end{table*}

The final yields for key isotopes are listed in Table \ref{table:yields}.
As expected, we find the explosion yields from the primary star to be similar to previous detonation studies of sub-Chandrasekhar WDs \citep{Fink_2010,Shen_2018_dets,Polin_2019,Gronow_2021,Boos_2021}.
Interestingly, the yields from our interior ignition companion detonations indicate more extensive burning than one would expect given the initial progenitor density.
This is demonstrated in the yields from the 1.00 $M_{\odot}$ WD across several models, where it has the same initial density profile.
When the 1.00 $M_{\odot}$ is detonated as the primary, or directly ignited as the companion, it produces 0.55 $M_{\odot}$ of $^{56}$Ni.
When it explodes as a companion after a interior ignition, however, it produces 0.61 $M_{\odot}$ of $^{56}$Ni.

\begin{figure}
    \centering
    \includegraphics[width=0.5\textwidth]{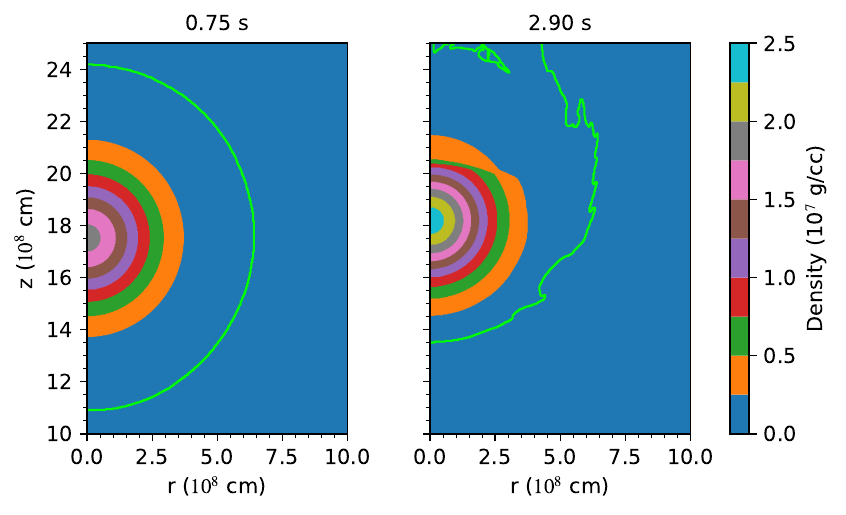}
    \caption{The qualitative density structure of a 0.90 $M_{\odot}$ companion before and after the impact of an exploding 1.00 $M_{\odot}$ primary. 
    The green line indicates the 90\% contour for companion material. 
    The first frame is just as the ejecta from the primary detonation is first reaching the companion while the second frame is at companion ignition time.
    This increase in density leads to enhanced burning in the companion detonation.
    }
    \label{fig:dens_enhancement}
\end{figure}

This increased production of radioactive material in the companion is due to the shock induced from the detonation of the primary, which increases the density in the bulk of the companion prior to ignition.
A demonstration of this density enhancement can be seen in Figure \ref{fig:dens_enhancement} which shows the density structure of the companion before impact and just before ignition.
While some of the outer star material has been spatially disrupted (indicated by the green contour lines in Figure \ref{fig:dens_enhancement}), the vast majority of the companion retains its spherical shape due to the low density enhancement relative to the original density of the inner WD.
This enhancement also creates a disagreement in the yields between the interior and direct ignition models of the 1.00 $M_{\odot}$ companion, as the initiation of the companion detonation in the direct ignition model coincides with the impact of the primary and thus the detonation propagates along the original density structure of the companion.

Some of our two star explosion models produce total amounts of $^{56}$Ni that fall within the range of that deduced from normal, observed SNe Ia, with our 1.00 + 0.90 $M_{\odot}$ model generating an amount near the expected upper limit  \citep{Stritzinger_2006,Scalzo_2014}.
Both of our 1.10 + 1.00 $M_{\odot}$ models, however, produce an amount of radioactive material that exceeds what is expected from normal SNe Ia and is more aligned with that of overluminous Type Ia events.

\subsection{Observables}
\label{subsec:synthetic_observables}
We split the presentation of our triple and quadruple detonation observables based on their ability to reproduce normal SNe Ia.
Our models that look like normal SNe Ia at and around maximum light are shown in Section \ref{subsubsec:snia_like_twostar} and our models that are overluminous, which arise from the 1.10 + 1.00 $M_{\odot}$ progenitor system, are shown in Section \ref{subsubsec:superch_cases}.
Additionally, we show the case where a companion does not detonate in Section \ref{subsubsec:1+.7_surviving}.

\subsubsection{SN Ia-like Two Star Explosions} 
\label{subsubsec:snia_like_twostar}

\begin{figure*}
    \centering    
    \includegraphics{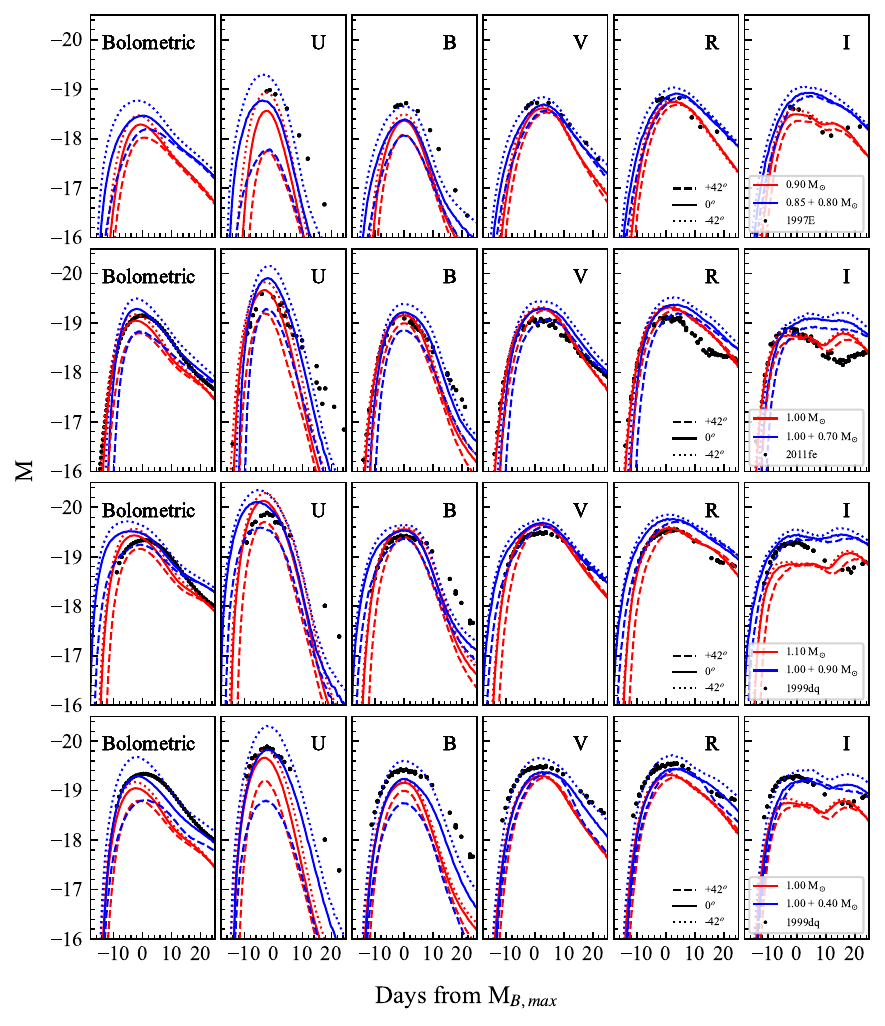}
    \caption{Multiband light curves of triple and quadruple detonation models that look SN Ia-like from this work (blue) compared with double detonation models from \citet{Shen_2021} (red).
    Three lines of sight from each model are shown, where the dotted, solid, and dashed lines represent each event as observed from a southern, equatorial, and northern line of sight, respectively.
    Also shown are light curves from observed SNe Ia with similar peak magnitudes.
    }
    \label{fig:lc_norms}
\end{figure*}

Light curves for our two star explosion models that have SN Ia-like observables are shown in Figure \ref{fig:lc_norms}.
Each of the models are compared to a single star, thin helium shell double detonation from \citet{Shen_2021} with a similar peak brightness.
We also show photometry for a few example observed SNe Ia: 1997E \citep{Hicken_2009}, 2011fe \citep{Munari_2013,Tsvetkov_2013}, and 1999dq \citep{Stritzinger_2006,Jha_2006,Ganeshalingam_2010}.
Light curves are corrected for Milky Way reddening as per \citet{Schlafly_2011}.

Overall, the shapes of the light curves of these triple and quadruple detonations are fairly similar to that of double detonations.
There are, however, some notable deviations between the one and two star detonations.
First, the rise times of the 0.85 + 0.80 $M_{\odot}$ and 1.00 + 0.90 $M_{\odot}$ are significantly longer than that of their one star counterparts, which is likely due to the increased amount of mass that is encapsulating both concentrations of radioactive ejecta (see Figure \ref{fig:primcmpn_and_ni56}).
This increase in rise time is more consistent with observation than single star double detonations or two star detonations where the companion is low mass (e.g. 1.00 + 0.70 $M_{\odot}$ and 1.00 + 0.40 $M_{\odot}$ models in Figure \ref{fig:primcmpn_and_ni56}) \citep{Yao_2019,Miller_2020,Fausnaugh_2023}.
An additional difference between these sets of models is an imbalance of agreement between different bands.
For example, the 0.85 + 0.80 $M_{\odot}$ model looks very similar to its counterpart in B, but much less so in bolometric and U.
Lastly, there are some disagreements between the models at different lines of sight.
This is most apparent in the 1.00 + 0.40 $M_{\odot}$ model where the northern line of sight shown is much less luminous in bolometric, U, and B than its counterpart, despite the agreement at the equatorial and southern lines of sight.
This can likely be attributed to the ashes of the low mass companion, which contain very little radioactive material, providing predominantly increased opacity at the line of sight that first intersects the companion ashes before that of the primary.
With companions of high enough mass ($\geq 0.80$ $M_{\odot}$), the companion ashes generate a significant amount of luminosity such that they do not have the luminosity deficit that the low mass companion models have at northern lines of sight.

An apparent delineator between double and quadruple detonations of similar brightnesses is the behavior of the I-band curve.
In most of the double detonations in Figure \ref{fig:lc_norms}, the I-band shows two distinct maxima: one just before the time of the B-band maximum and another $\sim$~20 days later.
In the quadruple detonation cases, however, the I-band remains fairly flat after rising to a peak and either shows a relatively weak secondary maximum or none at all.
This suppression of the secondary peak is also line of sight dependent, with the equatorial and southern viewing angles showing more monotonic evolutions of the I-band.

\begin{figure}
    \centering    \includegraphics[width=0.5\textwidth]{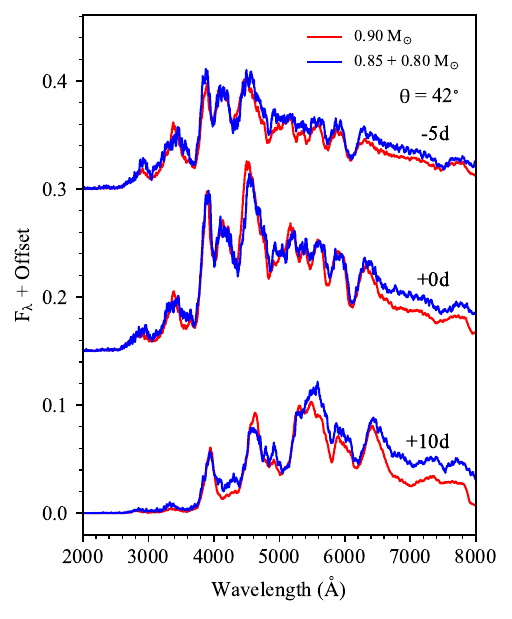}
    \caption{Spectral comparison between the 0.90 $M_{\odot}$ thin shell double detonation from \citet{Shen_2021} (red) and 0.85 + 0.80 $M_{\odot}$ quadruple detonation from this work (blue) at a viewing angle 42$^{\circ}$ above the equatorial plane.
    Each pair of spectra are at a different time relative to the B-band maxima.
    }
    \label{fig:m085_m080_spectra}
\end{figure}

\begin{figure}
    \centering
    \includegraphics[width=0.5\textwidth]{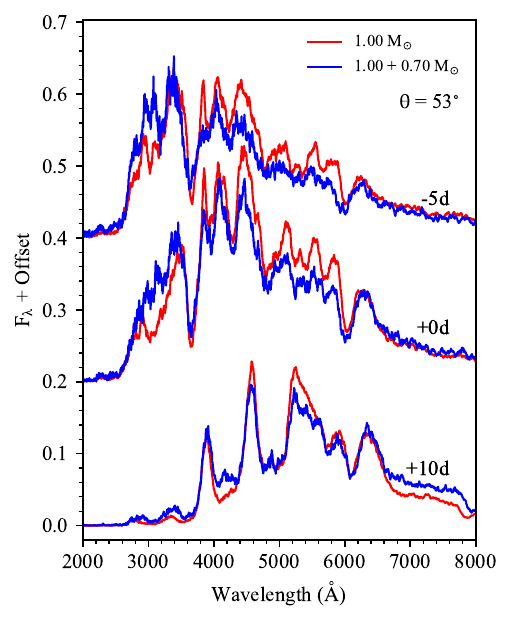}
    \caption{Like Figure \ref{fig:m085_m080_spectra}, but for a comparison between a 1.00 $M_{\odot}$ and 1.00 + 0.70 $M_{\odot}$ models at a viewing angle of 53$^{\circ}$.
    }
    \label{fig:m100_m070_spectra}
\end{figure}

\begin{figure}
    \centering    \includegraphics[width=0.5\textwidth]{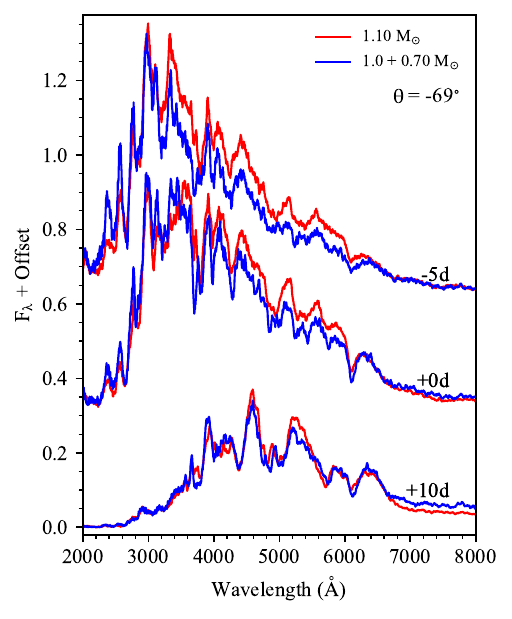}
    \caption{Like Figure \ref{fig:m085_m080_spectra}, but for a comparison between the 1.10 $M_{\odot}$ and 1.00 + 0.70 $M_{\odot}$ models at a viewing angle of -69$^{\circ}$.
    }    \label{fig:m100_m070_spectra_m110_alternate}
\end{figure}

\begin{figure}
    \centering    \includegraphics[width=0.5\textwidth]{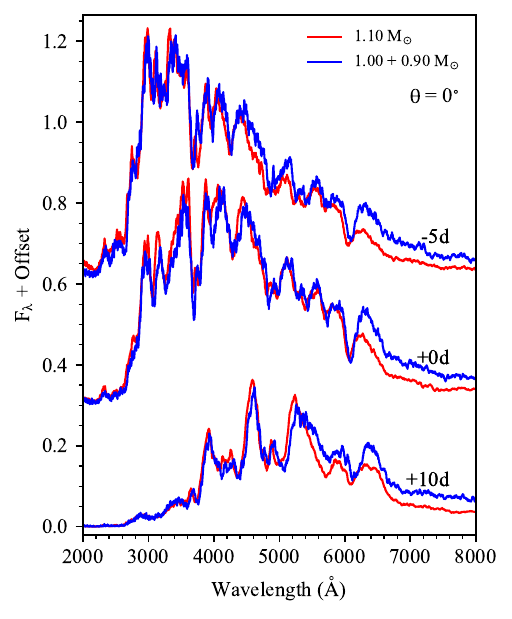}
    \caption{Like Figure \ref{fig:m085_m080_spectra}, but for a comparison between the 1.10 $M_{\odot}$ and 1.00 + 0.90 $M_{\odot}$ models at a viewing angle of 0$^{\circ}$.
    }
    \label{fig:m100_m090_spectra}
\end{figure}

\begin{figure}
    \centering    \includegraphics[width=0.5\textwidth]{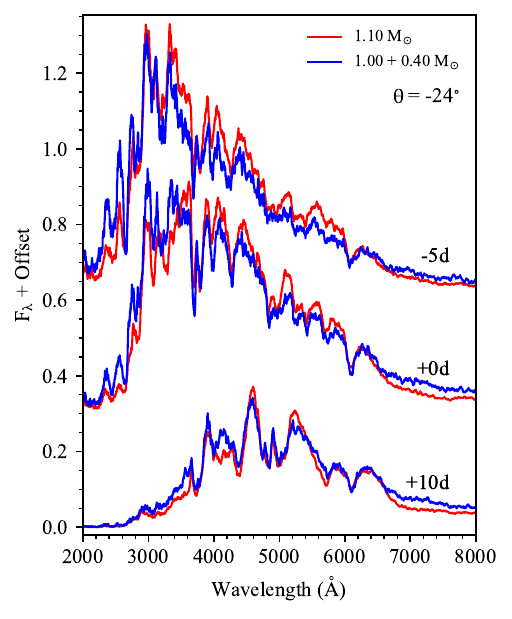}
    \caption{Like Figure \ref{fig:m085_m080_spectra}, but for a comparison between the 1.10 $M_{\odot}$ and 1.00 + 0.40 $M_{\odot}$ models at a viewing angle of -24$^{\circ}$.
    }    \label{fig:m100_m040_helium_companion_spectra}
\end{figure}

\begin{figure}
    \centering    \includegraphics[width=0.5\textwidth]{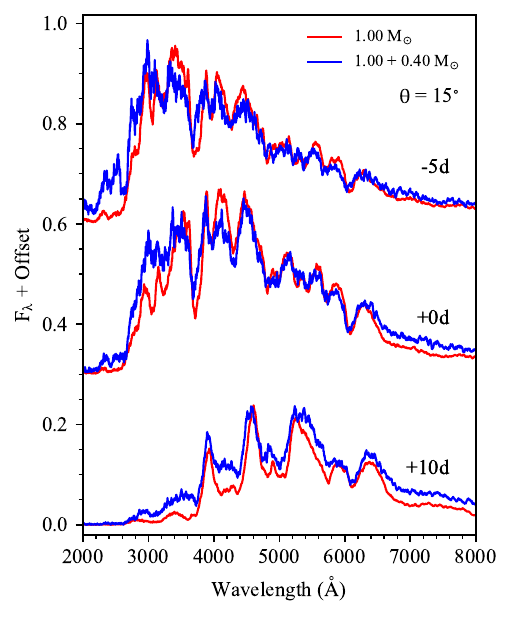}
    \caption{Like Figure \ref{fig:m085_m080_spectra}, but for a comparison between the 1.00 $M_{\odot}$ and 1.00 + 0.40 $M_{\odot}$ models at a viewing angle of 15$^{\circ}$.
    }    \label{fig:m100_m040_helium_companion_spectra_m100_alternate}
\end{figure}

Spectral comparisons for these models are shown in Figures \ref{fig:m085_m080_spectra}-\ref{fig:m100_m040_helium_companion_spectra} where they are compared to thin shell double detonation models with similar peak brightnesses from \citet{Shen_2021}.
For these progenitor systems with total masses below 2 $M_{\odot}$, we find that the triple and quadruple detonation models mimic the traditional double detonation remarkably well at this epoch, especially given the drastic difference in total ejecta mass.
In general, most spectral features and their characteristics that are seen in double detonation models and SN Ia observations are reproduced in these two star explosion models.
For example, in Figure \ref{fig:m085_m080_spectra}, much of the optical portion of the spectra before and at maximum light are nearly identical between the double and quadruple detonation models.
Similarities are seen throughout a range of progenitor system masses.

The triple and quadruple detonations do, however, have some variations compared to isolated double detonations.
One major difference is the enhanced UV in some of the two star models.
This is most prominently observed in Figure \ref{fig:m100_m070_spectra}, where the detonation of a 0.70 $M_{\odot}$ companion drives extra emission between 3000 and 4000 {\AA} around maximum light that is not observed with the sole detonation of the primary.
Differences between the one and two star detonation scenario can also be found in the IR.
This is demonstrated in Figure \ref{fig:m100_m090_spectra} where the spectra are nearly identical up to 6000 {\AA}, but the quadruple detonation case sees consistently higher emission beyond that wavelength.

We observe no clear helium signatures in the 1.00 + 0.40 $M_{\odot}$ triple detonation case (Figures \ref{fig:m100_m040_helium_companion_spectra}-\ref{fig:m100_m040_helium_companion_spectra_m100_alternate}) that leaves behind 0.16 $M_{\odot}$ of helium, primarily from the helium WD companion.
This is mostly due to the fact that the helium rich region of the ejecta is embedded at low velocities underneath the ashes of the primary and thus have little effect on the spectral features around maximum light.
We note that sub-Chandrasekhar detonations across a range of masses are expected to have a non-trivial amount of helium within their ejecta and that this may have an observational signature \citep{Collins_2023}.
However, the prediction of this signature requires the use of non-LTE.

\begin{figure}
    \centering \includegraphics[width=0.5\textwidth]{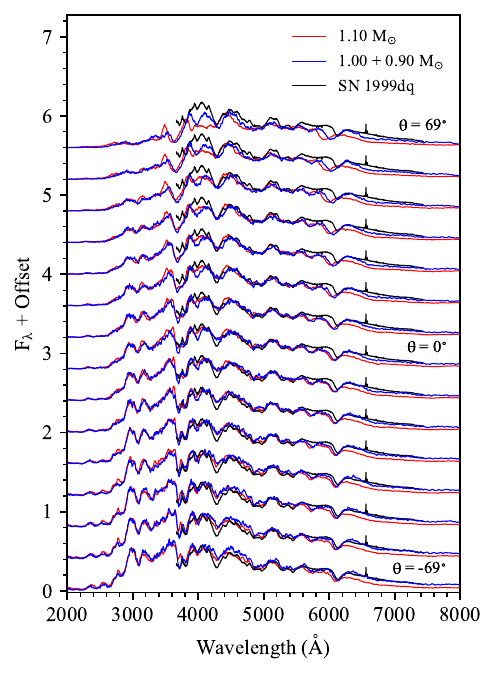}
    \caption{Spectra at all lines of sight for a 1.10 $M_{\odot}$ thin shell double detonation model from \citet{Shen_2021} and 1.00 + 0.90 $M_{\odot}$ quadruple detonation model in addition to SN 1999dq \citep{Matheson_2008}.
    Each spectrum is shown at +0.8 days from its respective maximum light time.
    }
    \label{fig:m100_m090_spectra_all_los}
\end{figure}

\begin{figure}
    \centering    \includegraphics[width=0.5\textwidth]{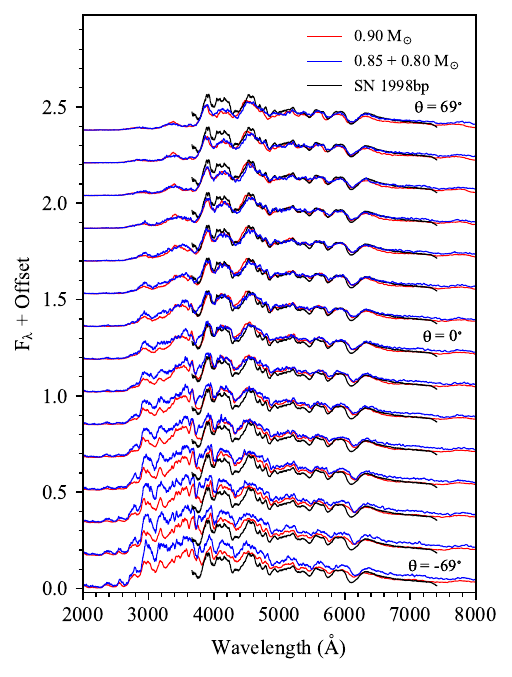}
    \caption{Like Figure \ref{fig:m100_m090_spectra}, but at -0.4 days from respective maximum light for the 0.90 $M_{\odot}$ and 0.85 + 0.80 $M_{\odot}$ models along with SN 1998bp \citep{Matheson_2008}.
    }
    \label{fig:m085_m080_spectra_all_los}
\end{figure}

In Figure \ref{fig:m100_m090_spectra_all_los}, we compare the spectra of double and quadruple detonation models at each line of sight at the times corresponding to their respective B-band maxima.
In this case, the spectra are very similar at most lines of sight up to \(\sim \)6000 {\AA}.
There is a slight decrease in Si\,{\sc ii}~$\lambda$6355\ velocity in the quadruple detonation model at most lines of sight, except at the northernmost lines of sight where the effect is significant.
However, this degree of consistency across viewing angles between the one and two star scenario does not hold across all mass configurations, which we exemplify in Figure \ref{fig:m085_m080_spectra_all_los}.
Interestingly, the lines of sight for this 0.85 + 0.80 $M_{\odot}$ quadruple detonation case are most dissimilar from the double detonation counterpart at southern angles where the primary intersects the observer's line of sight before the companion.
At these angles in some of our models, the spectra of the quadruple detonation models see increased emission between 4000 and 6000 {\AA} beyond maximum light (see B and V light curves for this model in Figure \ref{fig:lc_norms}).
This effect is possibly due to the compacted region of high-mass, radioactive material that is mostly concentrated in the southern regions of the ejecta (see Figure \ref{fig:primcmpn_and_ni56}).
Alternatively, the ejecta density and structure in the north of the quadruple detonation model is less of a departure from the double detonation, leading to more consistent observables from lines of sight in that direction.

Also shown in Figures \ref{fig:m100_m090_spectra_all_los} and \ref{fig:m085_m080_spectra_all_los} are comparisons with observed SNe Ia from \citet{Matheson_2008}.
These observed spectra, along with the following shown in this work, are scaled by a uniform factor at all lines of sight and dereddened for galactic extinction using the SNooPy Python tool for SNe Ia \citep{Burns_2011}, using the CCM prescription \citep{Cardelli_1989} and extinction values from \citet{Schlafly_2011}.
We find that the two star explosions models fit these observations of SNe Ia about as well as isolated double detonation models do.

\subsubsection{Overluminous SN Ia Candidate Models} \label{subsubsec:superch_cases}

\begin{figure}
    \centering    
    \includegraphics[width=0.5\textwidth]{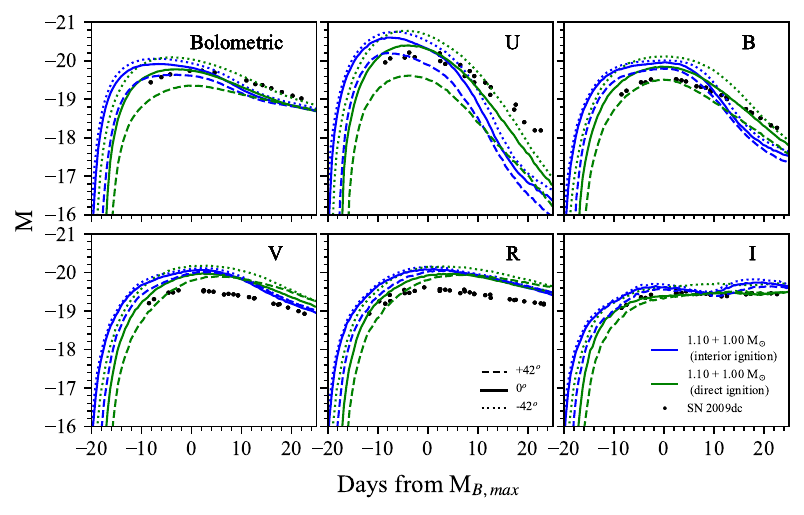}
    \caption{Multiband light curves from the two explosions modes of our most massive progenitor system, compared with overluminous SN Ia, SN 2009dc \citep{Silverman_2011,Taunenberger_2011}.
    }
    \label{fig:lc_superch}
\end{figure}

We present two models from our 1.10 + 1.00 $M_{\odot}$ progenitor system which differ based on the ignition mechanism of the companion; the ``interior ignition'' case, which is ignited within the core as the primary ejecta shock is passing through it (like most of the other models in this work), and the ``direct ignition'' case, which sees a natural ignition at the southern edge of the star that is coincident with the primary ejecta impact.
The light curves from these models are shown in Figure \ref{fig:lc_superch}.
We find that these two high-mass models reach much higher brightnesses and have significantly flatter light curve shapes post-peak than our previously shown models.
The interior and direct ignition cases differ in their light curves due to the different timing in the secondary detonations, and thus nucleosynthetic yields and ejecta structure, as detailed in Section \ref{subsec:yields}.
We compare these models in Figure \ref{fig:lc_superch} to observed overluminous SN Ia, SN 2009dc \citep{Silverman_2011,Taunenberger_2011}, which has relatively similar light curve shapes.

\begin{figure}
    \centering       \includegraphics[width=0.5\textwidth]{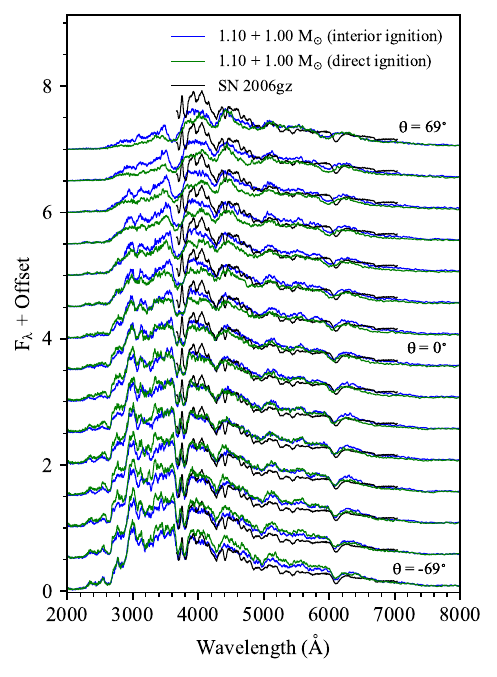}
    \caption{Spectra at all lines of sight from our two cases originating from a 1.10 + 1.00 $M_{\odot}$ binary system, compared with observed luminous SN Ia, SN 2006gz \citep{Hicken_2007}.
    Both model and observed spectra shown are from 2 days before their respective B-band maxima.
    }
    \label{fig:m110_m100_spectra_all_los}
\end{figure}

We also show the maximum light spectra at all lines of sight from our two 1.10 + 1.00 $M_{\odot}$ cases in Figure \ref{fig:m110_m100_spectra_all_los}, where they are compared with observed overluminous SN Ia, SN 2006gz \citep{Hicken_2007}.
Our two models show an interesting line of sight effect where they are fairly similar at equatorial and southern lines of sight, but show great differences at northern viewing angles.
We find a fairly decent match with SN 2006gz at southern lines of sight where there is strong agreement in overall spectral shape and the prominent features at 3800 and 4300 {\AA}.
A notable aspect in SN 2006gz that our models do not reproduce is the Si\,{\sc iii}~$\lambda$4560\ feature \citep{Hicken_2007}.
However, we will show in an upcoming work that the production of this line, among others, requires the use of non-LTE for our detonation models.
We also find significantly high Si\,{\sc ii}~$\lambda$6355\ velocities at northern viewing angles of these high-mass models, in disagreement with SN 2006gz and other overluminous SNe Ia \citep{Ashall_2021}.
These bright models also share many spectral similarities with the 91T-like class of overluminous SNe Ia, albeit for a much poorer reproduction of the Si\,{\sc ii}~$\lambda$6355\ feature.

While these models do not generate a complete match to observations of overluminous SNe Ia, they do indicate a potential path towards a mechanism that may produce such events which, like normal SNe Ia, still remain elusive in their origins.
A full consideration of the triple and quadruple detonation scenario leading to any class of overluminous SNe Ia would require a greater variety of examined binary configurations than just that from the pair of exploratory models shown in this work.

\subsubsection{1.00 + 0.70 $M_{\odot}$, One Star Explosion (Surviving Companion)} \label{subsubsec:1+.7_surviving}

\begin{figure}
    \centering    \includegraphics[width=0.5\textwidth]{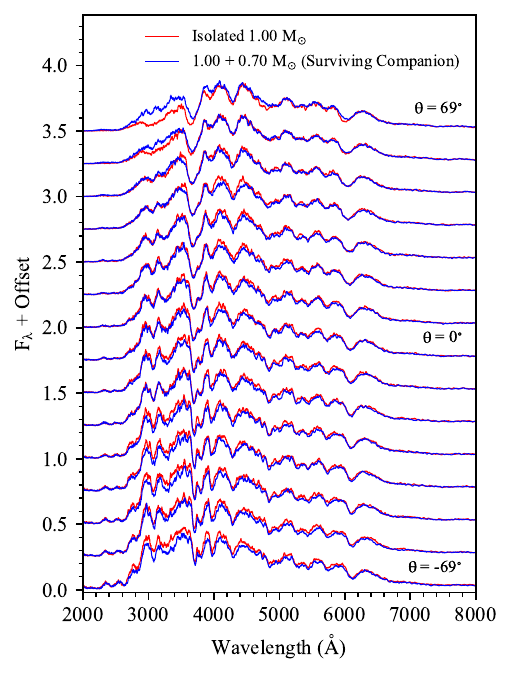}
    \caption{Spectra at all lines of sight from the detonation of a bare 1.0 $M_{\odot}$ WD with and without a companion.
    In the model where a companion is included, it is not ignited.
    }
    \label{fig:m100_m070_nosecdet_spectra_all_los}
\end{figure}

In Figure \ref{fig:m100_m070_nosecdet_spectra_all_los}, we show max light spectra from a 1.00 + 0.70 $M_{\odot}$ degenerate binary where only the primary detonates. 
We compare this to the detonation of an isolated, bare 1.00 $M_{\odot}$ WD, revealing the effects that the companion's presence has on the observables of such an event.
It is shown in Figure \ref{fig:m100_m070_nosecdet_spectra_all_los} that the presence of a 0.70 $M_{\odot}$ companion has little effect on the synthetic observables in our setup.
The biggest differences are at the northernmost line of sight, as one might expect, but the effect is mostly limited to the near-UV.
There is also a modest reduction of the Si\,{\sc ii}~$\lambda$6355\ line velocity ($\sim 2,000$ km s$^{-1}$) exclusive to this line of sight.
At mid and southern latitudes, the spectral features are nearly identical except for a slight decrease in overall luminosity.
This slightly decreased luminosity is perhaps attributable to the removal of the 0.04 $M_{\odot}$ of bound primary star ash, which is made up of mostly $^{56}$Ni.
While it is not considered in this work, the bound material may still effect the observables of the event.
For example, the decay of the bound radioactive isotopes may drive a wind, influencing late time observables \citep{Shen_2017}.
In contrast, 0.004 $M_{\odot}$ of companion material ends up entrained in the unbound ejecta in our simulation.

\subsection{Observational Correlations}
\label{subsec:observational_correlations}
One of the attractive features of the double detonation model shown in \citet{Shen_2021} is that it is possible to approximately span the observed range of SNe Ia parameters by varying just the mass of the exploding star and the line of sight. 
An obvious concern is that including another exploding star might adversely impact this feature.
Here we separately address the bolometric Phillips relation, the relations between peak magnitude, color and Si velocity, and, for the first time for our models, the gamma-ray escape time.

\subsubsection{Bolometric Phillips Relation}
\label{subsubsec:bolo_phillips}

\begin{figure}
    \centering    
    \includegraphics[width=0.5\textwidth]{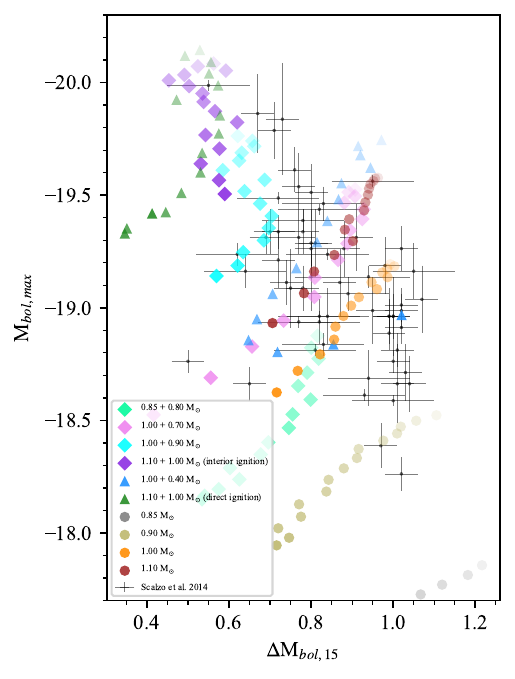}
    \caption{The bolometric Phillips relation for our two star explosion models from this work and a selection of isolated, thin helium shell double detonation models from \citet{Boos_2021,Shen_2021}.
    The new triple and quadruple detonation models are represented by triangles and diamonds, respectively, while double detonation models are represented by circles.
    The degree of transparency for each symbol denotes the line of sight, with the northernmost line of sight being the least transparent.
    The error bars are for observed normal SNe Ia from \citet{Scalzo_2014}.
    }
    \label{fig:philips_bolo}
\end{figure}

We show the bolometric Phillips relation for our new models in Figure \ref{fig:philips_bolo}, along with four thin shell, isolated double detonation models from \citet{Boos_2021}.
In general, the negative relationship between peak bolometric magnitude and decline rate, in addition to the scatter, is reproduced by our two sets of models.
When looking at an individual model across all viewing angles, however, the relationship between the two parameters is usually positive.
In the isolated double detonations, this trend is strictly positive and effectively linear, but some two star cases show departures from this.
For example, the 1.00 + 0.40 $M_{\odot}$ model has a sharp turn at the northernmost lines of sight. 
Additionally, the highest mass two star explosion models have a non-linear relationship between peak magnitude and decline rate as well as a narrower breadth of decline rates.

There are a couple regions of the observed bolometric parameter space that our relatively sparse model grid does not directly cover.
It is conceivable that models with progenitor masses between those shown here would likely reach that space.
For example, models with M$_{bol,max}$ $\sim-18.7$ and $\Delta$M$_{bol,15}$ $\sim1.0$ mag would likely arise from progenitors of $\sim0.95$ $M_{\odot}$.
Additionally, a 1.00 $M_{\odot}$ progenitor with an exploding companion that has a mass between 0.70 and 0.90 $M_{\odot}$ may fill in the high luminosity space at a $\Delta$M$_{bol,15}$ of $\sim0.75$ mag.

Some of our models also lay outside the observed parameter space in Figure \ref{fig:philips_bolo}.
The 1.10 + 1.00 $M_{\odot}$ models have decline rates that are modestly smaller than observed luminous events.
Our low mass cases, including the 0.85 and 0.90 $M_{\odot}$ isolated double detonations and most lines of sight from the 0.85 + 0.80 $M_{\odot}$ quadruple detonation, have peak magnitudes lower than what is observed in normal SNe Ia.
It may be possible that events that involve progenitors of this low mass produce peculiar SN Ia or do not actually detonate in reality.

\subsubsection{Peak Magnitude, Color, and Si\,{\sc ii} Velocity}
\label{subsubsec:peak_color_sivel}

\begin{figure*}
    \centering    
    \includegraphics{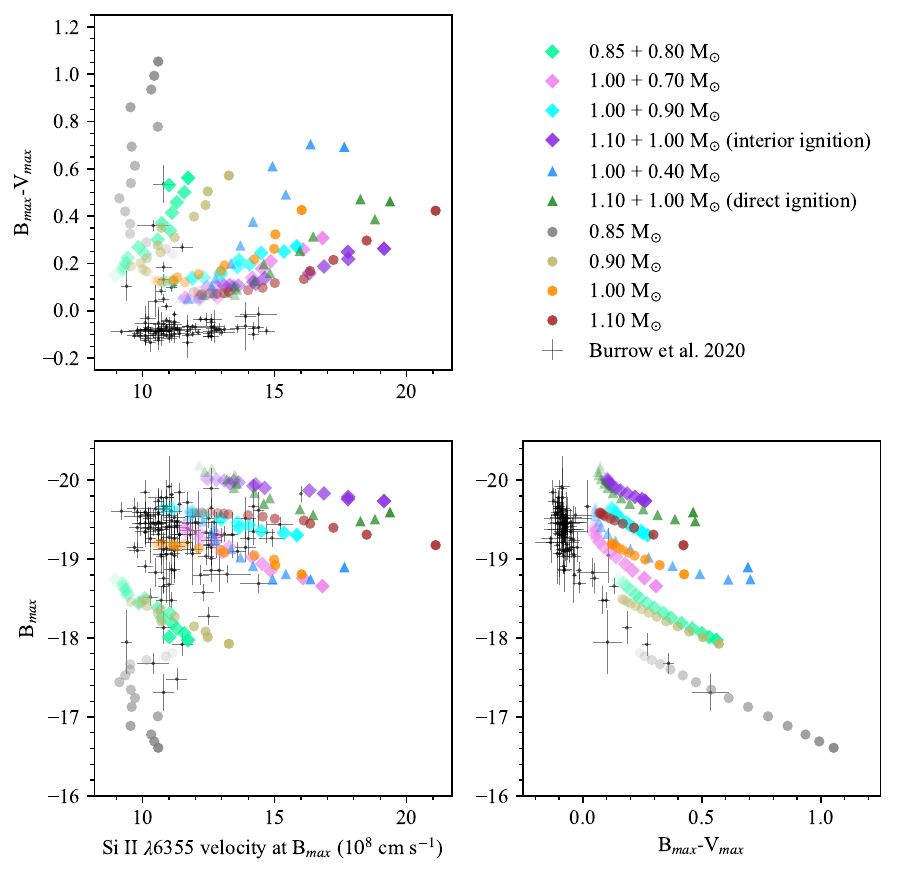}
    \caption{Observable correlations of models from this work and \citet{Boos_2021,Shen_2021} for Si\,{\sc ii}~$\lambda$6355\ velocity, B$_{max}$, and B$_{max}-$V$_{max}$.
    The symbols representing the models have the same scheme as Figure \ref{fig:philips_bolo}.
    The error bars are for observed SNe Ia from \citet{Burrow_2020}.
    }
    \label{fig:obs_characteristics}
\end{figure*}

An array of observational correlations for both one and two star explosion models is shown in Figure \ref{fig:obs_characteristics}.
We use the tool \texttt{Spextractor} ( \citet{Papadogiannakis_2019}, modified by \citet{Burrow_2020}) to calculate the photospheric velocities of our model spectra to be consistent with the methods from \citet{Burrow_2020}, from which the observational data points plotted in Figure \ref{fig:obs_characteristics} originate.
Overall, the two star explosion scenario below 2.0 $M_{\odot}$ resides generally within the extent of these observational parameters from the isolated double detonation.
The effects from the inclusion of the companion on the Si\,{\sc ii} velocities in the two star explosion models relative to their isolated double detonation counterparts depends on the mass of the companion.
For example, the 1.00 + 0.70 quadruple detonation $M_{\odot}$ model shows slightly higher velocities than the 1.00 $M_{\odot}$ double detonation case.
Alternatively, the 1.00 + 0.90 $M_{\odot}$ quadruple detonation model has a much slower velocity extent compared to its 1.10 $M_{\odot}$ counterpart.
The 1.00 + 0.40 $M_{\odot}$ triple detonation model shows particularly high Si\,{\sc ii}~$\lambda$6355\ velocities exclusively at the three northernmost viewing angles.
This is likely owed to the direct ignition mode that is unique to this and one of the 1.10 + 1.00 $M_{\odot}$ cases.
The 1.10 + 1.00 $M_{\odot}$ direct ignition model, however, does not see a significant velocity increase compared to the same progenitor system that underwent an interior ignition.

The effects on B$_{max}-$V$_{max}$ in the two star explosion scenario also depend on companion mass.
For example, the 0.85 + 0.80 and 0.90 $M_{\odot}$ models show significant consistency across B$_{max}-$V$_{max}$, while the 1.00 + 0.70 $M_{\odot}$ model shows bluer peak colors than both the 1.00 and 1.10 $M_{\odot}$ isolated double detonation cases.
Additionally, the two direct ignition cases see broader values for B$_{max}-$V$_{max}$ than their counterparts.

The double, triple, and quadruple detonation models presented in Figure \ref{fig:obs_characteristics} do a poor job at recreating the fairly narrow distribution of B$_{max}-$V$_{max}$ in observed SNe Ia.
This is at least partially due to our assumption of LTE in our radiative transfer calculations.
As discussed in \citet{Shen_2021} (see error bars in their Figure 20), all the parameters shown in Figure \ref{fig:obs_characteristics} are subject to non-LTE effects, with B$_{max}-$V$_{max}$ and Si\,{\sc ii}~$\lambda$6355\ velocity having significant potential non-LTE corrections.

\subsubsection{Gamma Ray Escape Time}
\label{subsubsec:gamma_escape}

\begin{figure}
    \centering    
    \includegraphics[width=0.5\textwidth]{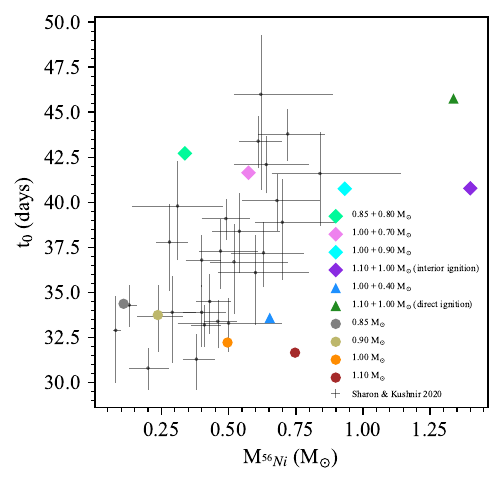}
    \caption{Estimated gamma ray escape times from one and two star explosion models from this work and \citet{Boos_2021}.
    The symbols representing the models have the same scheme as Figure \ref{fig:philips_bolo}.
    The error bars are for observed SNe Ia from \citet{Sharon_2020_data}.
    }
    \label{fig:gamma_escape}
\end{figure}

An aspect of SN Ia observables that the two star explosion scenario would affect is the gamma ray escape time.
In Figure \ref{fig:gamma_escape}, we show the gamma ray escape times and $^{56}$Ni yields for our previous, isolated double detonation models and new triple and quadruple detonation models, along with that from observed SNe Ia \citep{Sharon_2020_data}.
We calculate these escape times for our models by assuming one scattering for each escaping photon, as detailed by \citet{Wygoda_2019}, using 1D averaged profiles of our models.
We note that these escape times, in addition to the deduced $^{56}$Ni yields, likely have a line of sight dependence that is not examined here.
For our quadruple detonation cases, we observe consistently higher gamma ray escape times due to both the enhanced density of the primary ejecta and production of $^{56}$Ni yields in some our companion detonations.
Like the one comparison presented in \citet{Pakmor_2022}, we find that some events can have significantly different gamma ray escape times despite having fairly similar $^{56}$Ni yields (e.g. 0.85 + 0.80 $M_{\odot}$ vs. 0.90 $M_{\odot}$).
While the suite of models of this work fails to directly reproduce the trend observed in SNe Ia, they demonstrate the flexibility of the double detonation scenario to possibly fit such a relation.
This is important as all previous SN Ia models have been unable to produce the positive relationship between between gamma ray escape time and ejected $^{56}$Ni mass in observed events \citep{Sharon_2020,Kushnir_2020}.

\section{Discussion} \label{sec:discussion}

This paper demonstrates that triple and quadruple detonations occurring consecutively in a degenerate binary have similar synthetic observables to conventional one star SN Ia models, at least around maximum light. 
While these two star events may appear ``normal'' in many respects, there are some observational differentiators around maximum light.
For example, we show comparisons between one and two star explosion models that have strong similarities across most of the optical spectrum, but they often differ in the UV or beyond 6000 {\AA}.
Additionally, quadruple detonations with companions of masses 0.80 $M_{\odot}$ and above show significantly longer rise times.
Another differentiator is the less pronounced or lack of I-band secondary maximum in quadruple detonation models. 
\citet{Pessi_2022} finds a lack of this I-band ``kink'' in ${\sim}25\%$ of SNe Ia in their sample. 
The I-band is poorly reproduced in LTE \citep{shen_2021_nlte}, however, and a proper examination of this I-band kink and many other features require more physically complete radiative transfer calculations.

While the near-max-light observational signatures from the influence of a surviving companion are shown to be relatively minor in this work, observations at much later times may show more clear signs of a companion interaction \citep{Ferrand_2022}. 
There may be signatures to be found from the surviving companion, which is significantly altered from its original state before the binary disruption \citep{Tanikawa_2018}.
It may be possible that the peculiar features of the observed candidate runaway WDs from a double detonation scenario \citep{Shen_2018,ElBadry_2023} may be attributable to this ejecta-companion interaction.

Though idealized as a one-parameter family when characterized by \citet{Phillips_1993}, it steadily became clear that the scatter among normal SNe Ia is larger than could be attributed to photometric error or variation in extinction.
Physically, this introduces a need for, at minimum, a secondary parameter in addition to the $^{56}$Ni yield.
Recent work on double detonations has indicated line of sight as a secondary parameter within this model.
However, if two star explosions are common enough to contribute a significant fraction of the population, this introduces at least one additional parameter. 
Misalignment of the ignition point with the axis connecting the two stars, discussed more below, may also make line-of-sight a two parameter family.
With the work here showing that one and two star explosions overlap in typical observational characteristics, there is hope that the continued success of the double detonation model may finally allow the complex mixture of additional parameters to be teased out from within the SN Ia population. 
It also provides a physical reason why a simple single secondary parameter has been elusive.
Even based on work here it appears likely that both line of sight and companion mass are similarly important parameters, so that a single secondary parameter would not be a prediction of the double detonation model if the possible explosion of both stars is included.

It is perhaps useful to note that although Arnett's law \citep{Arnett_1982} is frequently used to infer brightness from $^{56}$Ni yield, that conversion requires the assumption or measurement of an ejecta velocity.
Thus a true one parameter family would only be generated if every $^{56}$Ni yield corresponded uniquely to a single ejecta velocity.
Both the line of sight dependence of ejecta velocity structure and the differences between the ejecta velocity structure of one and two star explosion cases seen here lead naturally to a family of velocities applicable to each $^{56}$Ni yield, thus appearing as additional parameters in the behavior of the light curve.
Additionally, real ejecta are, of course, more complex than the simple structure used in Arnett's law.

An original impetus of this work was the high Si\,{\sc ii}~$\lambda$6355\ velocities seen in our higher mass double detonation models from \citet{Shen_2021} at some lines of sight when compared to observation (see Figure \ref{fig:obs_characteristics}).
It was postulated that the presence of a surviving companion may slow down the ejecta material at northern lines of sight, where the highest photospheric velocities are seen.
As detailed in Section \ref{subsubsec:1+.7_surviving}, the photospheric velocity for the case of an exploding primary and surviving 0.70 $M_{\odot}$ companion is decreased slightly at the northernmost line of sight only.
Additionally, we show in Figure \ref{fig:obs_characteristics} that the Si\,{\sc ii}~$\lambda$6355\ velocities are slightly decreased for quadruple detonation models with companion masses of 0.85 and 0.90 $M_{\odot}$ relative to their one star explosion counterparts.
This is somewhat intuitive in our quadruple detonation setup given that the locations of the core ignitions in the primary and companion are located in opposite hemispheres (i.e. south for primary, north for companion).
This leads to detonations that are directed mostly towards each other (primary upwards, companion downwards), lowering the overall asymmetry of the ejecta.
The increase in brightness due to both stars making $^{56}$Ni also appears to improve the agreement with observation.
Brighter observed explosions that would previously have been attributed to higher individual masses (e.g. 1.10 $M_\odot$), that show too-high velocities along some lines of sight, might instead be attributed to lower-mass double explosions (e.g. 1.00 + 0.90 $M_\odot$) that do not have lines of sight that exhibit such high velocities.
While these effects do appear to make the theoretical high velocity outliers less of a problem, the overall distribution of velocities at most other lines of sight remain
generally higher than the observed distribution.
Overall, we find the issue of the unrealistically high Si\,{\sc ii}~$\lambda$6355\ velocities of our double detonation models to be unsolved.
The discrepancy of photospheric velocities between double detonation models and observation may be alleviated by a more careful treatment of the detonation (i.e. do our models produce $^{28}$Si out to too low of densities in the core detonation?) and/or radiative transfer. 

While this work has shown that the triple and quadruple detonation scenario may be observationally viable across a range of binary configurations, more work on the mechanics of two double detonations is needed in order to confidently determine if these events could occur in nature.
More precise calculations, particularly those that include the full system in 3D (like \citealt{Pakmor_2022}), may reveal dynamics of the detonation that are relevant to the resulting observables. 
Given that the helium detonation was not treated in this work, it is unclear to what extent the upward-moving shock from the primary core detonation (see Figure \ref{fig:dens_enhancement}) influences the ignition point of the companion core, which is normally triggered solely by the converging shock of its own helium shell.
If this factor leads to a shorter delay between core detonations, less primary ash would be able to wrap around the companion before it detonates (see Figure \ref{fig:primcmpn_and_ni56}). 
This would limit the amount of primary ash at northern latitudes, influencing the observables with a line of sight dependence.
Additionally, a more complete examination of this scenario could show how much helium is remaining on the surface of the companion at the time of the first ignition.
If it is shown, for example, that the companion has a very limited amount of helium on its surface for certain mass configurations, then the quadruple detonation may be impossible in those binaries.
The explosion modes in two star scenarios may also vary from those that are included in this work (e.g. a helium detonation that travels up the accretion stream, directly to the donor star; \citealt{pakmor_2021}).

Due to dimensional limitations, we have also only examined the scenario in which the core of the primary is ignited, within its southern hemisphere, along the orbital axis of the system.
This corresponds to a double detonation of the primary WD in which its helium shell is also ignited along the orbital axis (at the northern pole of the primary, nearest the companion).
This initial helium ignition is expected to be triggered near the impact point of the accretion stream which can have a notable separation from the orbital axis \citep[see Figure 1 in][]{Guillochon_2010}.
It is very likely that the location of initial helium ignition in the primary would lead to noticeable differences in the formation of the ejecta and observables, if not a major change in detonation dynamics.
For example, a helium detonation that begins at the equator (90$^{\circ}$ from the implicit and explicit helium ignition locations of this work and \citealt{Boos_2021}, respectively) would ultimately lead to primary ejecta that are directed more strongly away from the companion than those in this work.
This would allow the fastest material in the ejecta to expand more freely as opposed to being slowed and compacted by the companion as in this work.
It may also affect the timing delay between WD detonations or even the occurrence of a companion detonation.
The robustness of the shock-induced ignition of the companion helium shell (or helium core) relative to the shock strength and binary separation is generally unaddressed by this work, but will be explored in detail by an upcoming related work.
We note that with cylindrical symmetry, it is not possible to realistically model the scenario where the helium ignition occurs anywhere other than a pole of the primary that is aligned with the central axis of the companion.
This work is also idealized in terms of the spherical symmetry of our progenitors at the time of detonation; in reality, the companion and its atmosphere would be substantially deformed by this point which would likely affect its ignition and detonation.

Another limitation of this work is our somewhat limited time domain of the synthetic observables, which are calculated out to around 50 days. 
We expect there to be much greater differences in the observables between the one and two star explosion scenario at later times when the photosphere moves inwards and reveals more information about the structure at lower velocities.
This is especially true for our case involving a detonating 0.40 $M_{\odot}$ helium companion that leaves behind 0.16 $M_{\odot}$ of unburnt He.
In addition to the nebular phase observables, the two star explosion scenario should also impact the long term shape of the supernova remnant.

While observed SN Ia remnants are considered to be fairly symmetric \citep{Lopez_2011}, double detonation ejecta are inherently somewhat non-spherical due to the off-centered ignition of the core.
The addition of a second doubly-detonating star, as in this work, affects the asymmetry in a complex manner.
The innermost ejecta in the quadruple detonation scenario is now much more spherical due to the companion material expansion being suppressed by the presence of relatively dense primary ashes surrounding it.
However, the outer ejecta, which consists of the primary star ashes, is now more aspherical primarily due to the companion exploding off-center to it.
It is not directly obvious how a supernova remnant may appear many years after the explosion based on its early ejecta morphology, but recent theoretical work involving a double detonation in a double degenerate system has shown that the original structure of the ejecta can have a lasting impact \citep{Ferrand_2022}.
Additionally, the previously described assumptions in our setup, in particular the imposed cylindrical symmetry of the calculations, may be leading to an overestimation of the final ejecta asymmetry.

A notable result of the two star explosion scenario compared to isolated sub-Chandrasekhar detonation models is the non-monotonic stratification of core ejecta abundances, which would likely lead to more significant differences in the nebular phase.
This is especially interesting in the context of unexplained double-peaked Fe lines observed in the the nebular spectra of some normal SNe Ia \citep{Dong_2015}.
This feature is indicative of a bimodal distribution of Fe in the supernova ejecta, which is found to have a peak separation of around 5,000 km s$^{-1}$ \citep{Dong_2015}.
This is interpreted as the result of a bimodal distribution of $^{56}$Ni generated in the explosion, which would decay to $^{56}$Fe by the nebular phase.
As suggested by \citet{Pakmor_2022}, it may be possible that the quadruple detonation scenario can lead to this conspicuous observational feature.
While the models shown in this work are not specifically consistent with the bimodal features detailed in \citet{Dong_2015}, they demonstrate a method to produce bimodal distributions of high-mass material along a line of sight (see Figure \ref{fig:all_ejecta}) following a SN Ia that may be interpreted as normal at maximum light.
It may be that alternative ignition locations (as described above) or lower mass companions (see \citealt{Pakmor_2022}) could produce this bimodal distribution of $^{56}$Fe that is consistent with some SNe Ia.

\section{Conclusion} \label{sec:conclusions}
In our 2D simulations we have determined that two WDs detonating subsequently can generate observables that are remarkably similar to double detonations of a single thin helium-shelled progenitor.
Given the work of the past several years that has shown the double detonation as a plausible Type Ia mechanism candidate, this new result suggests that the complete destruction of a WD binary via the triple or quadruple detonation may also be a Type Ia channel, provided the appropriate mass configuration.

Given this, there are several avenues for future work that may elucidate whether these events could occur in nature and how they may be differentiated from other Type Ia explosion mechanisms, including a double detonation of a single star.
More ``full picture'' simulations, like those performed in \citet{Tanikawa_2019,Pakmor_2022}, could more precisely establish where and when the companion ignition occurs, in addition to exploring higher-dimensional effects that are not accessible in this work.
While we showed that the detonation of a thin helium shell has little observational effect in the isolated double detonation scenario, it is possible that the shell ejecta may play a larger role in the quadruple explosion scenario, particularly for the companion shell ejecta that would be compressed.
A wider parameter space of progenitor characteristics, including other binary mass configurations and alternative metallicities, would also serve our understanding of the two star explosion scenario and its possible contribution to the scatter of observed SNe Ia.
Lastly, the use of non-LTE radiative transfer calculations would improve our interpretation with observation, most notably allowing for an examination of how the two star explosion scenario may affect the Phillips relation \citep{Phillips_1993,shen_2021_nlte}.

The full post-processed yields, ejecta profiles, and synthetic spectra out to ${\sim}$50 days for the new detonation models shown in this work can be found on Zenodo (\href{https://doi.org/10.5281/zenodo.10515767}{10.5281/zenodo.10515767}).

\begin{acknowledgments}
We thank the referee for their helpful comments. S.J.B. acknowledges support from NASA grant HST AR-16156. Financial support for K.J.S. was in part provided by NASA/ESA Hubble Space Telescope programs \#15871 and \#15918.

Resources supporting this work were provided by the NASA High-End Computing (HEC) Program through the NASA Advanced Supercomputing (NAS) Division at Ames Research Center and the University of Alabama high performance computing facility.

\software{\texttt{FLASH} (\citealt{Fryxelletal2000,Dubeyetal2009,Dubeyetal2013,Dubeyetal2014}, \url{flash.uchicago.edu}),
\texttt{MESA} (\citealt{paxton_2011,paxton_2013,paxton_2015,paxton_2018}, \url{mesa.sourceforge.net}),
\texttt{Sedona} (\citealt{Kasen_2006}),
\texttt{yt} (\url{yt-project.org}),
\texttt{matplotlib} (\citealt{matplotlib}, \url{matplotlib.org}),
\texttt{SNooPy} (\citealt{Burns_2011}),
\texttt{Spextractor} (\url{github.com/anthonyburrow/spextractor}).
}

\end{acknowledgments}

\bibliography{ref}{}
\bibliographystyle{aasjournal}



\end{document}